\documentclass[submission, Proceedings]{SciPost}

\hypersetup{
    colorlinks,
    linkcolor={red!50!black},
    citecolor={blue!50!black},
    urlcolor={blue!80!black}
}

\usepackage[bitstream-charter]{mathdesign}
\usepackage{graphicx}
\usepackage{epstopdf}
\usepackage{bm}
\usepackage{color}
\usepackage{soul}
\usepackage{mathtools}
\usepackage{xfrac}
\usepackage{rotating}
\usepackage{enumitem}
\usepackage{scrextend}
\usepackage{slashed}
\usepackage{subfig}
\usepackage{braket}

\DeclareSymbolFont{usualmathcal}{OMS}{cmsy}{m}{n}
\DeclareSymbolFontAlphabet{\mathcal}{usualmathcal}

\newcommand\Eqn[1]     {Eq.\,(\ref{#1})}

\newcommand\eqn[1]     {eq.\,(\ref{#1})}
\newcommand\eqns[2]    {eqs.\,(\ref{#1}) and~(\ref{#2})}

\def\beq{\begin{equation}}
	\def\eeq{\end{equation}}
\def\beqa{\begin{eqnarray}}
	\def\eeqa{\end{eqnarray}}

\def\nn{\nonumber}

\newcommand{\al}{\alpha}

\newcommand{\mm}{\mathcal{M}}
\newcommand{\eps}{\epsilon}
\newcommand{\tts}{\mathbf{T}_t^2}
\newcommand{\tsu}{\mathbf{T}_{s-u}^2}

\newcommand{\mTree}{\mathcal{M}^{\text{tree}}}
\newcommand{\ag}[1]{\hat{\alpha}_g^{(#1)}}

\newcommand{\mExpM}[2]{\hat{\mathcal{M}}^{(-,#1,#2)}}

\newcommand{\T}{{\bf T}}

\def\Tt{{\bf T}_t^2} 
\def\Tsu{{\bf T}_{s{-}u}^2}

\allowdisplaybreaks

\begin{document}

\begin{center}{\Large \textbf{
Two-parton scattering in the high-energy limit: \\ 
climbing two- and three-Reggeon ladders }}
\end{center}

\begin{center}
G. Falcioni\textsuperscript{1},
E. Gardi\textsuperscript{1},
N. Maher\textsuperscript{1},
C. Milloy\textsuperscript{2},
L. Vernazza\textsuperscript{2,3$\star$}
\end{center}

\begin{center}
{\bf 1} Higgs Centre for Theoretical Physics, School of Physics and Astronomy,\\ The University of Edinburgh, Edinburgh EH9 3FD, Scotland, UK
\\
{\bf 2} Dipartimento di Fisica and Arnold-Regge Center, Universit\'{a} di Torino, \\
and INFN, Sezione di Torino, Via P. Giuria 1, I-10125 Torino, Italy
\\
{\bf 3} Theoretical Physics Department, CERN, Geneva, Switzerland
\\
* Leonardo.Vernazza@to.infn.it
\end{center}

\begin{center}
\today
\end{center}


\definecolor{palegray}{gray}{0.95}
\begin{center}
\colorbox{palegray}{
  \begin{tabular}{rr}
  \begin{minipage}{0.1\textwidth}
    \includegraphics[width=35mm]{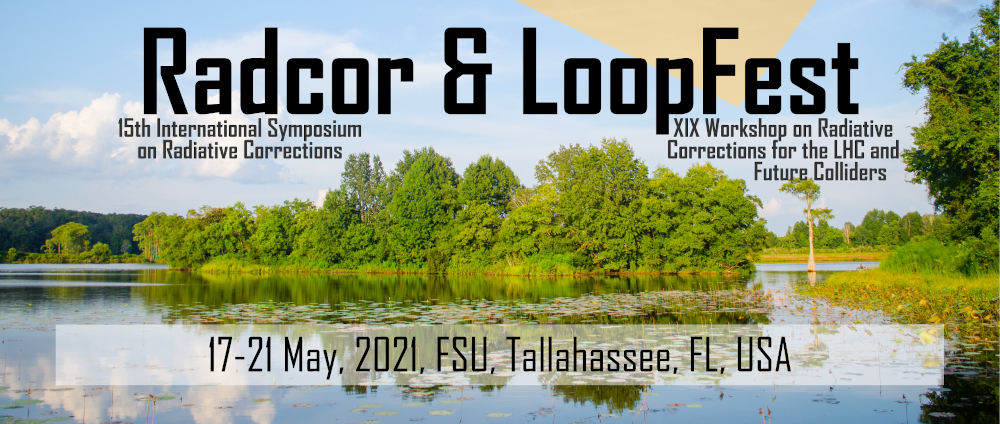}
  \end{minipage}
  &
  \begin{minipage}{0.85\textwidth}
    \begin{center}
    {\it 15th International Symposium on Radiative Corrections: \\Applications of Quantum Field Theory to Phenomenology,}\\
    {\it FSU, Tallahasse, FL, USA, 17-21 May 2021} \\
    \doi{10.21468/SciPostPhysProc.?}\\
    \end{center}
  \end{minipage}
\end{tabular}
}
\end{center}

\section*{Abstract}
{\bf
We review recent progress on the calculation of scattering amplitudes 
in the high-energy limit. We start by illustrating the shockwave formalism, 
which allows one to calculate amplitudes as iterated solutions of rapidity 
evolution equations. We then focus on our recent results regarding $2\to 2$ 
parton scattering. We present the calculation of the imaginary part of the 
amplitude, at next-to-leading logarithmic accuracy in the high-energy 
logarithms, formally to all orders, and in practice to 13 loops. We then 
discuss the computation of the real part of the amplitude at
next-to-next-to-leading logarithmic accuracy and through four loops. 
Both computations are carried in full colour, and provide new 
insight into the analytic structure of scattering amplitudes and their 
infrared singularity structure.}

\vspace{5pt}
\noindent\rule{\textwidth}{1pt}
\tableofcontents\thispagestyle{fancy}
\noindent\rule{\textwidth}{1pt}
\vspace{5pt}

\section{Introduction}
\label{sec:intro}

The high-energy, or Regge limit of gauge-theory scattering amplitudes
offers a unique insight into gauge dynamics. In this limit amplitudes 
simplify and factorize in rapidity, revealing new degrees of freedom 
that propagate in two transverse dimensions. Rapidity evolution equations, 
such as BFKL \cite{Kuraev:1977fs,Balitsky:1978ic} and its non-linear generalization~\cite{Balitsky:1995ub,Kovchegov:1999yj,JalilianMarian:1996xn,JalilianMarian:1997gr,Iancu:2001ad},  
allow us to translate concepts from Regge theory~\cite{Collins:1977jy}
into calculation tools, leading to concrete predictions for partonic amplitudes \cite{Dixon:2012yy,Dixon:2014voa,Caron-Huot:2013fea,Caron-Huot:2017fxr,Caron-Huot:2017zfo,Caron-Huot:2020grv,Gardi:2019pmk,DelDuca:2019tur,Bartels:2020twc}. The study of the Regge 
limit has been crucial in determining multi-leg planar $ \mathcal{N} $ = 4 
super Yang-Mills (SYM) amplitudes in general kinematics to unprecedented 
accuracy, see e.g.  \cite{Dixon:2014voa,Caron-Huot:2016owq,Caron-Huot:2019vjl,DelDuca:2019tur}. 
In parallel, computation of the Regge limit in $2\to 2$ scattering for general colour~\cite{Caron-Huot:2013fea,Caron-Huot:2017fxr,Caron-Huot:2017zfo,Caron-Huot:2020grv} 
have been shown to provide powerful constraints on soft singularities of 
amplitudes in general kinematics~\cite{DelDuca:2011ae,Henn:2016jdu,Almelid:2017qju,Gardi:2019pmk}.

In this talk we focus on $2\to 2$ gauge-theory scattering amplitudes, 
and review recent results in \cite{Caron-Huot:2017zfo,Caron-Huot:2020grv,Falcioni:2020lvv,upcoming}, obtained within a framework developed in~\cite{Caron-Huot:2013fea,Caron-Huot:2017fxr}. 
In a scattering configuration $1+2 \to 3+4$, with Mandelstam invariants 
$s\equiv (p_1+p_2)^2$, $t\equiv (p_1-p_4)^2$ and 
$u\equiv (p_1-p_3)^2=-s-t$, the high-energy limit is defined by the condition 
$s\gg -t$. Assuming that the momentum transfer $-t$ is large compared to the 
QCD scale, these amplitudes can be calculated as an expansion in the strong 
coupling, and notoriously develop large logarithms in the ratio $\frac{s}{-t}$. 
It has long been known that the Leading Logarithms (LLs) are resummed 
to all orders~\cite{Lipatov:1976zz,Kuraev:1976ge} via
\begin{equation}
\mm^{\text{LL}}_{ij\to ij}(s,t) 
=  \left(\frac{s}{-t}\right)^{C_A \,\al_g(t)}\!{\cal M}^{\rm tree}_{ij\to ij},
\qquad {\rm with} \qquad
{\cal M}^{\rm tree}_{ij\to ij}= g_s^2\frac{2s}{t} T_i \cdot T_j, 
\label{eq:ReggeTraj}
\end{equation}
where the colour generator $T_i$ is in the representation of parton~$i$, and
\begin{equation} \label{al_g}
\al_g = \frac{\al_s}{\pi}\frac{r_\Gamma}{2\eps}+{\cal O}(\al_s^2)\,;
\qquad 
r_\Gamma = e^{\eps\gamma_E}\frac{\Gamma^2(1-\eps)\Gamma(1+\eps)}{\Gamma(1-2\eps)}
\end{equation} 
is the gluon Regge trajectory, written in dimensional regularization 
with $D=4-2\epsilon$. Higher-order corrections, not written in \eqn{al_g}, 
contribute beyond~LL. The simple exponentiation property in \eqn{eq:ReggeTraj} 
can be interpreted as due to the exchange of an effective degree of freedom, 
which is identified as a single \emph{Reggeized gluon}, or \emph{Reggeon}. 
At higher logarithmic accuracy the scattering amplitude develops a more 
complex analytic structure, which can be understood in term of the exchange 
of multiple Reggeons~\cite{Low:1975sv,Nussinov:1975mw,Gunion:1976iy} in the 
transverse plane. These involve evolution equations which are integrable in 
the planar limit~\cite{Lipatov:1993yb,Faddeev:1994zg,Derkachov:2001yn,Derkachov:2002wz}, 
but are difficult to solve in general. However, in the pertubative regime 
they can be integrated iteratively~\cite{Caron-Huot:2013fea,Caron-Huot:2017fxr,Caron-Huot:2017zfo,Caron-Huot:2020grv}, 
to obtain perturbative high-energy amplitudes order-by-order in $\alpha_s$.

The factorization structure of the amplitude beyond LLs becomes clearer 
by exploiting the symmetry under the exchange $s \leftrightarrow u$, 
also known as \emph{``signature''} symmetry. to this end, the 
scattering amplitude is split into odd ($-$) and even ($+$) component 
with respect to $s \leftrightarrow u$:   
\beq
\mm = \,\,\mm^{(-)} + \mm^{(+)}.
\eeq
It is also convenient to expand the amplitude into a signature-symmetric 
definition for the large logarithm:
\begin{equation}\label{L_def}
\frac12\left(\log\frac{-s-i0}{-t}+\log\frac{-u-i0}{-t}\right) 
= \log\left|\frac{s}{t}\right| -i\frac{\pi}{2}\equiv L \,,
\end{equation}
such that one has
\begin{equation}\label{eq:expansionDef}
{\mm}^{(\pm)}_{ij\to ij} = \sum_{n=0}^\infty \left(\frac{\al_s}{\pi}\right)^n \sum_{m=0}^nL^m{\mm}^{(\pm,n,m)}_{ij\to ij}, \qquad 
{\rm with} \qquad 
{\mm}^{(-,0,0)}_{ij\to ij}\equiv {\cal M}_{ij\to ij}^{\rm tree}.
\end{equation} 
Splitting the amplitudes into $(\pm)$ components offers several advantages: 
first of all, signature is preserved under rapidity evolution, which 
greatly simplifies the computation of these amplitudes~\cite{Caron-Huot:2013fea,Caron-Huot:2017fxr,Caron-Huot:2017zfo,Caron-Huot:2020grv}. 
Furthermore, given that a single Reggeon has the quantum numbers of a 
gluon exchanged in the $t$ channel, which is odd under the signature 
symmetry, one finds that $\mm^{(-)}$ consists of an odd number of 
Reggeons, while $\mm^{(+)}$ involves an even number of them~\cite{Caron-Huot:2013fea}. 
In addition, there are other useful properties: for instance, it 
is possible to show~\cite{Caron-Huot:2017fxr} that the odd amplitude 
coefficients $\mm^{(-,n,m)}$ are purely \emph{real}, while 
the even ones $\mm^{(+,n,m)}$ are purely \emph{imaginary}, when 
expanded in powers of the symmetric logarithm $L$. 
Moreover, for gluon-gluon and quark-gluon scattering Bose symmetry 
links the kinematic dependence to that of colour, therefore 
$\mm^{(+)}$ and $\mm^{(-)}$ are governed by $t$-channel exchange 
of colour representations which are respectively even and odd 
under $1\leftrightarrow 4$ (or $2\leftrightarrow 3$) interchange. 

\begin{figure}[t]
	\centering
	\includegraphics[width=0.50\textwidth]{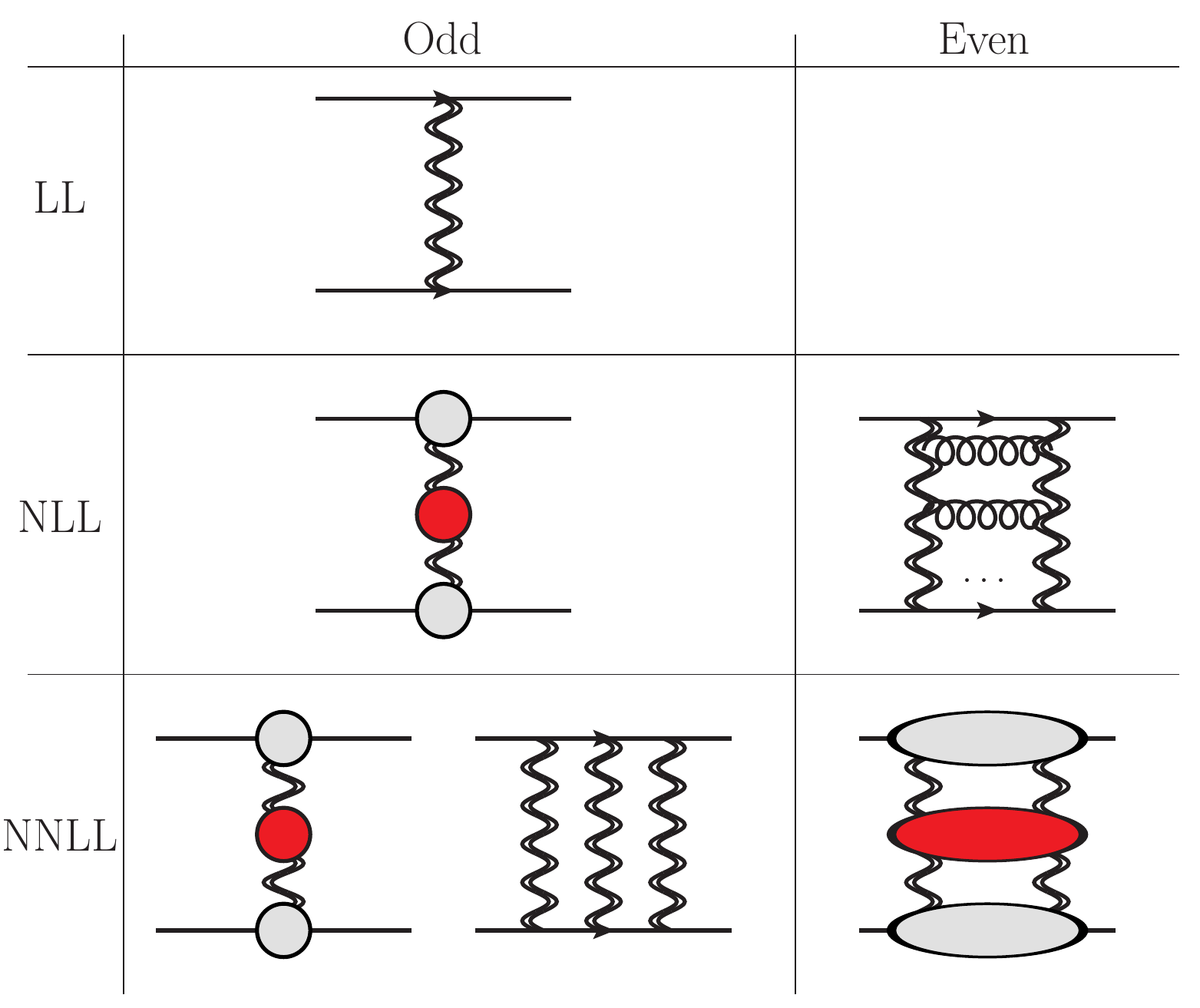}
	\caption{Schematic relation between logarithmic counting and Reggeon exchanges for two-parton amplitudes in the high-energy limit. The grey blobs represent impact factors, while the red ones are corrections to the Regge trajectory.}
	\label{fig:fact}
\end{figure}
One obtains a schematic representation for the amplitude, which 
connects the logarithmic accuracy with the number of Reggeon exchanges, 
as represented in fig. \ref{fig:fact}. One has 
\beq \label{polecut}
{\cal M}^{(-)} = {\cal M}^{(-),\,\rm SRS} + {\cal M}^{(-),\,\rm MRS}\,, 
\qquad \qquad
{\cal M}^{(+)} =  {\cal M}^{(+),\,\rm MRS}\,
\eeq
where the superscripts labels amplitudes involving single-Reggeon 
states (SRS) and multi-Reggeon states (MRS), respectively. As shown 
in fig. \ref{fig:fact}, the odd component $\mm^{(-)}$ 
is given up to NLL accuracy by a single Reggeon exchange~\cite{Fadin:2006bj,Fadin:2015zea}, 
with ${\cal O}(\alpha_s^2)$ corrections to the trajectory in 
(\ref{al_g}), along with $s$-independent impact factors~\cite{Fadin:1995xg,Fadin:1996tb,Fadin:1995km,Blumlein:1998ib}:
\begin{equation}
	\label{Regge-Pole-General}
	\mm^{(-),\,\text{LL+NLL}}_{ij\to ij} =
	\mm^{(-),\, \rm SRS}_{ij\to ij}\big|_{\rm NLL} = 
	e^{C_A \alpha_g(t) L} 
	C_i(t) \, C_j(t)
	{\cal M}^{\rm tree}_{ij\to ij}\,.
\end{equation}
The factors $C_{i}(t)$ are expanded in perturbation theory, 
with coefficients $C^{(n)}_i$. The $n = 1$ term contributes at 
NLL, while the higher-order terms give rise to further subleading 
logarithms. 

A proper description of the scattering amplitude beyond 
\eqn{Regge-Pole-General} requires to take into account the 
$t$-channel exchange of multi-Reggeon states. The signature 
even amplitude $\mm^{(+)}$, which starts at NLL accuracy, 
is governed at this logarithmic order by two-Reggeon exchange \cite{Caron-Huot:2013fea,Caron-Huot:2017zfo}. 
Furthermore, starting with the odd amplitude $\mm^{(-)}$ 
at NNLL accuracy, one needs to take into account the 
exchange of three Reggeons and their mixing with 
one-Reggeon states~\cite{Caron-Huot:2017fxr,Fadin:2016wso}. 
In this talk we review recent results obtained in~\cite{Caron-Huot:2017zfo,Caron-Huot:2020grv} 
and~\cite{Falcioni:2020lvv,upcoming}, where the two towers of 
coefficients ${\mm}^{(+,n,n-1)}$ and ${\mm}^{(-,n,n-2)}$ have 
been expressed to all orders as iterated solution of the BFKL 
equation and its generalization, the Balitsky-JIMWLK equation. 
Thanks to the development of new methods, the 
${\mm}^{(+,n,n-1)}$ tower has been evaluated explicitly 
to 13 loops,~\cite{Caron-Huot:2020grv}, while the tower 
${\mm}^{(-,n,n-2)}$ has been computed to 4 
loops in~\cite{Falcioni:2020lvv,upcoming}.

The calculation of scattering amplitudes in terms of Reggeon 
exchanges is particularly interesting, because it allows one 
to connect modern perturbation theory with concepts such as 
Regge poles and cuts, obtaining a deeper understanding of the 
analytic structure of scattering amplitudes. In general, an 
amplitude that can be brought to the form of eq.~(\ref{Regge-Pole-General}) 
is said to admit \emph{Regge-pole factorization}. It is only 
admitted by the signature odd amplitude, to NLL accuracy.
The signature even amplitude is described in turn by a Regge 
cut associated with a two-Reggeon exchange. Starting at NNLL 
accuracy, also the signature odd amplitude receives 
contributions due to multi-Reggeon states, giving 
rise to Regge cuts, such that one has
\begin{equation}
	\label{Schemes_MRS}
	{\cal M}_{ij\to ij}^{(-),\,\text{LL+NLL+NNLL}} =  e^{C_A \alpha_g(t) L} \,
	C_i(t) \, C_j(t) \,
	{\cal M}^{\rm tree}_{ij\to ij} 
	\,+ \,{\cal M}_{ij\to ij}^{(-),\,\rm MRS}\,.
\end{equation}
The second term in the r.h.s. first arises at NNLL at two 
loops~\cite{DelDuca:2001gu,DelDuca:2014cya}, and it has been 
interpreted as being due to three-Reggeon exchange in refs.~\cite{Caron-Huot:2017fxr,Fadin:2016wso,Fadin:2017nka}, 
where the three-loop NNLL amplitude was determined. As a 
consequence of \eqn{Schemes_MRS}, at NNLL accuracy $C_{i/j}(t)$ 
and $\al_g(t)$ start to depend on how the separation in 
eq.~(\ref{polecut}) is precisely defined, which is 
referred to below as a scheme choice. Following \cite{Caron-Huot:2017fxr}, 
we adopt the scheme in which the Regge cut contribution 
is then identified with the MRS, which is explicitly computed using the 
Balitsky-JIMWLK rapidity evolution equation. The Regge-pole
contribution is identified with the SRS contribution, determined by 
matching the r.h.s of \eqn{Schemes_MRS} to the full amplitude. Other 
schemes may be possible, and we refer to \cite{upcoming} for a more 
detailed discussion on this topic. 

An additional motivation to investigate multi-loop corrections in the 
high-energy limit is that it provides information on the infrared 
singularity structure of amplitudes. Further discussion of this 
aspect can be found in the original papers \cite{DelDuca:2011ae,Caron-Huot:2013fea,Caron-Huot:2017fxr,Caron-Huot:2017zfo,Falcioni:2020lvv,upcoming} and in a talk given at this conference \cite{Maher:2021nlo}, 
to which we refer for further details.

\section{Methodology}
\label{sec:methods}

Following~\cite{Caron-Huot:2013fea,Caron-Huot:2017fxr}, we describe 
two parton scattering in the high-energy limit within the shockwave 
formalism: fast particles moving in the $(+)$ lightcone direction, 
defined as \emph{projectile}, scatter against fast particles moving 
in the $(-)$ lightcone direction, or \emph{target}. The fast 
projectile $\ket{\psi_i}$ appears as a set of infinite Wilson lines 
$U(z_1) \otimes \cdots \otimes U(z_n)$~\cite{Korchemskaya:1994qp,Korchemskaya:1996je} 
at transverse position $z_k \equiv x_{k\perp}$, crossing the target 
$\bra{\psi_j}$ (seen as a ``shockwave'') at $x_- = 0$, with
\beq
\label{Udef}
U(z) = \mathcal{P}\exp \left[ig_s \mathbf{T}^a\!\!
\int_{-\infty}^{+\infty}\!\!\!dx^+ A^a_+(x^+,x^-\!=\!0,z) \right].
\eeq
In perturbation theory the unitary matrices $U(z)$ are close to the 
identity, therefore they can be parameterized in terms of a 
colour-adjoint field $W^a$:
\beq
U(z) = e^{ig_s\,T^aW^a(z)}\,, \label{Uparam}
\eeq
which is identified as a source for Reggeized gluons. The projectile 
and target are thus expanded in Reggeon fields: schematically, 
\beq
\label{sum_of_n_Reggeons}
\ket{\psi_i} =\sum_{n=1}^{\infty}\ket{\psi_{i,n}}
\sim g_s\, \big[\T \, W\big]_i(p) - g_s^2 \big[ \T \, W \otimes \T \,W\big]_i(p)
- g_s^3 \big[\T\, W \otimes \T \,W\otimes \T\, W \big]_i(p)+ \ldots,
\eeq
where $\ket{\psi_{i,n}}$ represents a state of $n$ Reggeons, 
which depends on the transverse momentum $p$, but not on the 
center-of-mass energy. The energy dependence enters through 
the fact that infinite Wilson lines develop rapidity 
divergences, regulated introducing a rapidity cutoff $\eta=L$, leading to 
a rapidity evolution equation for the projectile (and the target)
\beq \label{rapidity_evolution}
-\frac{d}{d\eta}\,\ket{\psi_i} = H\, \ket{\psi_i},
\eeq
where $H$ is the Balitsky-JIMWLK Hamiltonian~\cite{Balitsky:1995ub,Kovchegov:1999yj,JalilianMarian:1996xn,JalilianMarian:1997gr,Iancu:2001ad}. 
A key feature of eq.~\eqref{rapidity_evolution} is the non-linearity 
of $H$: evolution of the full projectile $\ket{\psi_i}$ in rapidity 
generates an increasing number of Wilson lines $U(z_j)$, eventually 
leading to the phenomenon of gluon saturation. However, in applications 
to partonic scattering processes it is appropriate to take the limit 
of dilute projectile and target~\cite{Caron-Huot:2013fea,Caron-Huot:2017fxr},
in which case $H$ acts on states with a given number of Reggeon 
fields~$W$. In this perturbative regime, 
$H$ is diagonal to leading order in $g_s^2$; the non-linearity of 
$H$ manifests at higher orders in the coupling, producing transitions 
between states with different number of Reggeon fields: a transition 
$k \to k+2n$ is of order $g_s^{2(1+n)}$. Also, note that, as a 
consequence of the signature symmetry, only transitions 
$k \to k \pm 2$ are allowed, while transitions $k \to k \pm 1$ 
are forbidden. In short, the r.h.s of 
eq.~\eqref{rapidity_evolution} has the form
{\small \begin{align} \label{Hamiltonian_schematic_form}
H  \left( \! \! \! \!
\begin{array}{c}
	W      \\  WW     \\  WWW  \\  \cdots
\end{array}
\!\! \! \! \right) &\equiv
\left( \!\!\!
\begin{array}{cccc}
	H_{1{\to}1} & 0  & H_{3{\to}1} & \ldots \\
	0 & H_{2{\to}2}  & 0  &  \ldots \\
	H_{1{\to}3} & 0  & H_{3{\to}3} & \ldots\\
	\cdots & \cdots  & \cdots & \cdots\\
\end{array} \!\!\!
\right)\left( \!\!\! \!
\begin{array}{c}
	W      \\  WW     \\  WWW  \\  \cdots\end{array}
\!\!\!\!\right) \simeq
\left( \!\!\!
\begin{array}{cccc}
	g_s^2 & 0  & g_s^4 & \ldots \\
	0 & g_s^2  & 0  &  \ldots \\
	g_s^4 & 0  & g_s^2 & \ldots\\
	\cdots & \cdots  & \cdots & \cdots\\
\end{array} \!\!\!
\right) \left( \!\!\! \!
\begin{array}{c}
	W      \\  WW     \\  WWW  \\  \cdots\end{array}
\!\!\! \!\right),
\end{align}}
\!\!where non-vanishing entries on the r.h.s. display 
the perturbative orders at which the various transition 
Hamiltonian start contributing. While off-diagonal entries 
are obtained from expansion of the Leading Order (LO) 
Balitsky-JIMWLK equation, diagonal entries are given 
by the Regge trajectory for $H_{1\to 1}$, or 
the LO BFKL equation for $H_{n \to n}$, with $n \neq 1$. 

A scattering amplitude in the high-energy limit is then given 
as an expectation value between states with fixed number of 
Reggeon fields $W$ evolved to equal rapidity:
\beq\label{eq:ampEqualsPsi}
\frac{i(Z_iZ_j)^{-1}}{2s}\mathcal{M}_{ij\to ij} 
= \braket{\psi_j|e^{-HL}|\psi_i}, 
\quad {\rm with}\quad 
Z_i(t)=\exp\left\{-\frac{1}{2}\int_0^{\mu^2}
\frac{d\lambda^2}{\lambda^2}\Gamma_i\left(\al_s(\lambda^2),\lambda^2\right)\right\}.
\eeq
Notice that, for simplicity, we define the expectation 
value in terms of states $\ket{\psi_i}$ where collinear 
divergences have been removed. These are given by the 
factors $Z_i(t)$, defined as an integral over the scale of 
$\Gamma_i$, where, to three loops, 
$\Gamma_i = \frac{1}{2}\gamma_K\left(\alpha_s(\lambda^2)\right)
C_i\log\frac{-t}{\lambda^2}+2\gamma_i$. Here $\gamma_K$ is the 
component of the cusp anomalous dimension~\cite{Korchemsky:1987wg,Gardi:2009qi,Gardi:2009zv,Becher:2009cu,Becher:2019avh}
proportional to the quadratic Casimir $C_i$, in the representation 
of parton $i$, and $\gamma_i$ are anomalous dimensions associated 
with on-shell form factors~\cite{FormFactors,DelDuca:2014cya}.

Our aim is to calculate transitions which involve multi-Regge states, 
therefore, it proves useful to introduce a \emph{reduced amplitude}, 
obtained from the original amplitude by removing single Reggeon transitions: 
\beq \label{eq:MhatDef}
\frac{i}{2s}\hat{\mm}_{ij\to ij} = e^{-\,\T_t^2\, \alpha_g(t) \, L}
\braket{\psi_j|e^{-H L}|\psi_i} \equiv 
\braket{\psi_j|e^{-\hat H L}|\psi_i},
\quad 
{\hat{H}}_{k\to k+2n} = H_{k\to k+2n} +\delta_{n0}\tts\al_g(t),
\eeq
which we have expressed in terms of a \emph{reduced} Hamiltonian 
${\hat{H}}$. After evolution has been performed, the contraction 
of Reggeons of equal rapidity is evaluated in terms of free propagators $\bra{W^a(p)}W^b(q)\rangle=\frac{i}{p^2}\delta^{ab}\delta^{2-2\epsilon}(p-q)+{\cal{O}}(g_s^2)$, see~\cite{Caron-Huot:2013fea}. We are now ready to discuss the 
calculation of the amplitudes ${\hat \mm}^{(+,n,n-1)}$ and 
${\hat \mm}^{(-,n,n-2)}$.

\section{Two-Reggeon cut}
\label{sec:twoReggeonCut}

\begin{figure}[t]
\begin{center}
\includegraphics[width=0.42\textwidth]{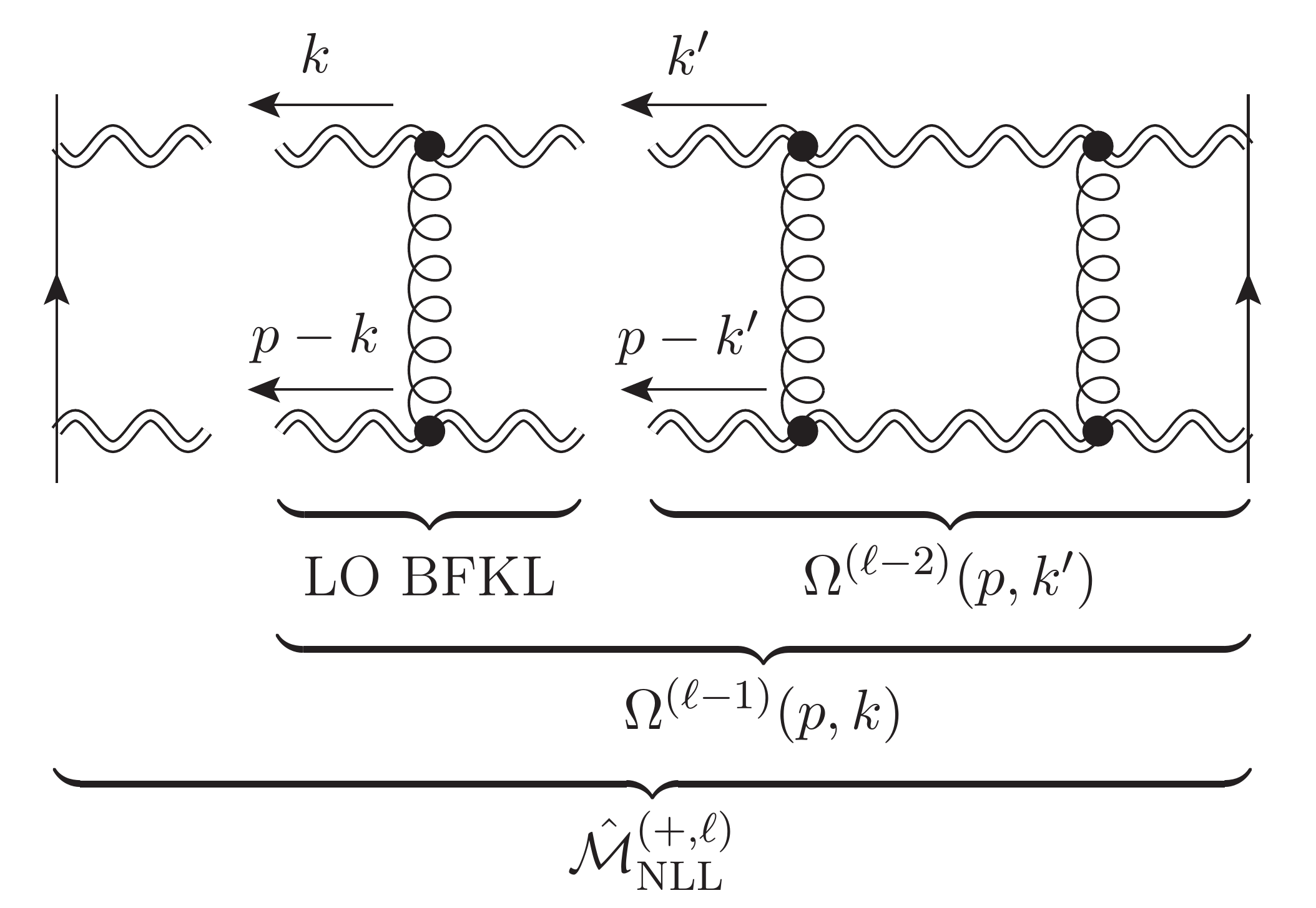}
\caption{Representation of the two Reggeon amplitude and the corresponding BFKL wavefunction. A single application of the BFKL Hamiltonian adds a rung in the ladder. The amplitude at $\ell$ loops, ${\cal M}^{(\ell)}(p)$, is obtained by integrating over the wavefunction $\Omega^{(\ell-1)}(p,k)$, which itself consists of $\ell-1$ rungs.}
\label{BFKL_Hamiltonian}
\end{center}
\end{figure}

According to \eqn{eq:MhatDef} and fig. \ref{fig:fact}, 
the NLL contribution to the even amplitude is given by 
\beq
\frac{i}{2s}{\hat \mm}_{ij\to ij}^{(+),\rm NLL} = \braket{\psi_{j,2}|e^{-\hat H L}|\psi_{i,2}},
\eeq
where the subscript of $| \psi_{i,2}\rangle$ identifies 
the two-Reggeon component of the projectile $i$ (and target $j$). 
Upon expansion one obtains \cite{Kuraev:1977fs,Balitsky:1978ic,Lipatov:1985uk,Caron-Huot:2013fea,Caron-Huot:2017zfo}:
\begin{align}
\label{Amp}
\hat{{\cal M}}_{ij\to ij}^{(+,\ell,\ell-1)}= 
-i\pi \frac{r_\Gamma^\ell}{(\ell-1)!} 
\int [Dk] \frac{p^2}{k^2(p-k)^2} \Omega^{(\ell-1)}(p,k) \, \Tsu \,
{\cal M}^{\rm tree}_{ij\to ij}\,,
\end{align}
where $[Dk] \equiv \frac{\pi}{r_\Gamma} 
\left( \frac{\mu^2}{4\pi e^{-\gamma_E}} \right)^{\epsilon} 
\frac{d^{2-2\eps} k}{(2\pi)^{2-2\eps}}$, and $r_\Gamma$ has been defined in \eqn{al_g}.
We refer to $\Omega^{(\ell-1)}(p,k)$ as the two Reggeized 
gluon wavefunction at $\ell-1$ loop order, as shown in 
fig.~\ref{BFKL_Hamiltonian}.
The wavefunction admits the BFKL evolution equation
\begin{equation}
\label{evol}
\frac{d}{dL}\, \Omega(p,k)=
\frac{\alpha_s r_{\Gamma}}{\pi} \, \hat{H} \, \Omega(p,k)\,,
\end{equation}
where the Hamiltonian has two components $\hat{H} = (2C_A-\Tt) \,\hat H_{\rm i} + (C_A-\Tt) \, \hat H_{\rm m}$ with
\begin{subequations}
\label{Him}
\begin{align}
\hat H_{\rm i}\,\Psi(p,k) 
=\int [{\mathrm D}k'] \, f(p,k,k') \Big[\Psi(p,k')-\Psi(p,k)\Big]\,,\quad\,\, 
\hat H_{\rm m}\,\Psi(p,k) = J(p,k)\, \Psi(p,k)\,,
\end{align}
\end{subequations}
where
\begin{equation}
\label{fk}
f(p,k,k') \equiv \frac{k^2}{k'^2(k - k')^2}
+\frac{(p-k)^2}{(p-k')^2(k - k')^2}
-\frac{p^2}{k'^2(p- k')^2}\,,
\end{equation}
and
\begin{align}\label{J}
J(p,k) &= \frac{1}{2\epsilon} + \int [{\mathrm D}k'] \, f(p,k,k') = \frac{1}{2\epsilon} \left[2- \left(\frac{p^2}{k^2}\right)^{\epsilon}-\left(\frac{p^2}{(p-k)^2}\right)^{\epsilon}\right].
\end{align}
While the exact solution is unknown, the equation can be solved iteratively 
\begin{equation}
\Omega(p,k) = \sum_{\ell=1}^{\infty} 
\left( \frac{\alpha_s\, r_{\Gamma}}{\pi}\right)^{\ell}  
\frac{L^{\ell-1}}{(\ell -1)!} \, \Omega{(\ell-1)}(p,k),\qquad \Omega^{(\ell-1)}(p,k) =
	\hat{H} \,\Omega^{(\ell-2)}(p,k),
\end{equation}
and we get
\begin{align} 
\Omega^{(0)}(p,k) &= 1, 
\qquad \qquad \qquad 
\Omega^{(1)}(p,k) = (C_A-{\mathbf T}_t^2)  J(p,k), \\ \nn
\Omega^{(2)}(p,k) &= (C_A-{\mathbf T}_t^2)^2 J^2(p,k)
+ (2C_A-{\mathbf T}_t^2) (C_A-{\mathbf T}_t^2) \! 
\int \! [{\mathrm D}k'] f(p,k,k') \left[ J(p,k') - J(p,k) \right]\,,
\end{align}
and so on. While for the first few orders it is possible to integrate directly these wavefunctions to get the dimensionally-regularized amplitude, this task becomes quickly intricate at higher orders. It is thus necessary to resort to more sophisticated techniques \cite{Caron-Huot:2017zfo,Caron-Huot:2020grv}.

There are two crucial observations that allows one to carry the integration in \eqns{Amp}{Him} to high order: the first is that the wavefunction $\Omega^{(\ell)}(p,k)$ is finite, to any order, which follows from the form of the Hamiltonian (\ref{Him}). As a consequence, all infrared singularities in the NLL amplitude must be generated in the final integration over $k$ in eq.~(\ref{Amp}), from the soft regions where $k\ll p$ or where $p-k\ll p$. The symmetry of the problem implies that these regions give the same contribution, hence it is sufficient to focus on the former, and multiply the result by two. The second observation is that BFKL evolution closes in the soft approximation, $k\ll p$. This is easily seen by considering a single evolution step of the wavefunction generated by the Hamiltonian (\ref{Him}), and observing that $f(p,k,k')$ vanishes in regions other than the soft region $k\sim k'\ll p$. Applying these arguments iteratively, we deduce that the soft limit of the wavefunction (and thus the complete singular part of the amplitude) is fully determined by the configuration where the entire side rail of the ladder is soft. It is thus possible to expand eq.~(\ref{Him}) and solve it for $k\sim k'\ll p$. This is much easier, because in this approximation the original two-scale problem, $k^2$ and $(p-k)^2$, is reduced to a single-scale problem. The integrals are then simple to perform, and the solution for the wavefunction at any order is given as a polynomial in $\xi \equiv (p^2/k^2)^\epsilon$:
\begin{align}
	\label{OmegaSoft}
	\Omega^{(\ell-1)}(p,k) =
	\frac{(C_A-{\mathbf T}_t)^{\ell-1}}{(2\epsilon)^{\ell-1}} \sum_{n=0}^{\ell-1} 
	(-1)^n 
	\left(\begin{array}{c}{\ell-1}\\{n}\end{array} \right)
	\left(\frac{p^2}{k^2}\right)^{n\epsilon} 
	\prod_{m=0}^{n-1} \left\{1 - \hat{B}_{m}(\epsilon) \frac{2C_A-{\mathbf T}_t}{C_A-{\mathbf T}_t}\right\}\,,
\end{align}
where $\hat B_n(\epsilon) = 2 n (2 + n) \zeta_3 \epsilon^3 + 3 n (2 + n) \zeta_4 \epsilon^4 +\ldots$. Performing the final integration over $k$ according to (\ref{Amp}) we obtain all the infrared-singular contributions to the NLL amplitude at any loop order $\ell$. Remarkably, the result can be resummed into a closed-form expression:
\begin{align}
\label{softAmplRes}
\left.\hat{\cal M}_{\rm NLL}^{(+)}\right|_s 
= \frac{i\pi}{L(C_A-{\mathbf T}^2_t)} 
\bigg( 1 - R(\epsilon) \frac{C_A}{C_A -{\mathbf T}_t^2} \bigg)^{-1} 
\left[ \exp\left\{\frac{r_{\Gamma}}{2\epsilon} 
\frac{\alpha_s}{\pi} L (C_A-{\mathbf T}_t)\right\} - 1 \right] 
{\mathbf T}^2_{s-u} \, {\cal M}^{({\rm tree})} 
\end{align}
up to  ${\cal O}(\epsilon^0)$ terms, where  
\begin{equation}\label{Reps}
R(\epsilon) 
=\frac{\Gamma^{3}(1-\epsilon)\Gamma(1+\epsilon)}{\Gamma(1-2\epsilon)} -1 
= -2\zeta_3 \, \epsilon^3 -3\zeta_4 \, \epsilon^4 -6\zeta_5 \epsilon^5
-\left(10 \zeta_6-2\zeta^2_3 \right) \epsilon^6 + {\cal O}(\epsilon^7).
\end{equation}

Now that the singularities have been determined, the remaining task it to compute the ${\cal O}(\epsilon^0)$ finite terms in the amplitude. Once again, the fact that the wavefunction is finite plays a central role~\cite{Caron-Huot:2020grv}: it implies that the wavefunction can be computed consistently in two transverse dimensions, setting $\epsilon=0$. This may seem to pose a problem for the subsequent calculation of the amplitude, given that the integral in (\ref{Amp}) requires (dimensional) regularization. However, this issue can be overcome by carefully combining the soft limit discussed above with the two-dimensional calculation~\cite{Caron-Huot:2020grv}.

The calculation of the wavefunction in exactly two transverse dimension is rather technical, and we refer to \cite{Caron-Huot:2020grv} for a detailed discussion. Let us mention here that this computation is done by representing the two-dimensional momentum $k$, $k'$ and~$p$ in the BFKL kernel of \eqn{Him} as complex parameters, 
$k = k_x + i k_y,$ \, $k' = k_x' + i k_y'$ and $p = p_x + i p_y$,
and implementing a variable change in the BFKL equation:
\begin{equation}
\label{eq:zwdef} \frac{k_x + i k_y}{p_x + i p_y} = 
\frac{z}{z-1} \qquad \text{and} \qquad \frac{k_x' + i k_y'}{p_x + i p_y} 
= \frac{w}{w-1}.
\end{equation}
It is possible to show that in these variables, the 2-dimensional wavefunction can be expressed in terms of single-valued harmonic polylogarithms (SVHPLs,~\cite{Brown:2004ugm,Brown:2013gia,Schnetz:2013hqa,Pennington:2012zj,Dixon:2012yy,DelDuca:2013lma}) at any order. Single valuedness is to be expected here, because branch cuts are physically inadmissible in the Euclidean two-dimensional transverse space. Moreover, the structure of the Hamiltonian guarantees that the wavefunction at order $\ell$ is a pure function of uniform weight $\ell$.
It is then possible to express the action of the 2-dimensional Hamiltonian on any linear combination of SVHPLs ${\cal L}_w(z,\bar z )$ (where the word $w$ corresponds to a set of 0 and 1 indices) as a set of differential equations, which can be integrated iteratively, using the soft limit $z,\bar z \to 0$ as boundary data. The procedure has been automated in~\cite{Caron-Huot:2020grv}. For instance, the first two orders reads:
\begin{align}
\label{Omega2d}
\Omega_{\rm 2d}^{(1)} &= 
\frac{1}{2} C_2 \left({\cal L}_0+2 {\cal L}_1\right), \\ \nn
\Omega_{\rm 2d}^{(2)} &= 
\frac{1}{2} C_2^2 \left({\cal L}_{0,0}+2 {\cal L}_{0,1}
+2 {\cal L}_{1,0}+4 {\cal L}_{1,1}\right)
+\frac{1}{4} C_1 C_2 \left(-{\cal L}_{0,1}
-{\cal L}_{1,0}-2 {\cal L}_{1,1}\right).
\end{align}

With the amplitude determined to all orders in the soft approximation, 
see \eqn{softAmplRes}, and the wavefunction in two dimension available to any required order, we return to the question of how to combine this information to obtain the full amplitude. This is  achieved~\cite{Caron-Huot:2020grv} by 
writing the full wavefunction as the sum of a soft and an hard components: $\Omega(p,k) = \Omega_\text{soft}(p,k) + \Omega_\text{hard}(p,k)$, where the two components are defined as follows: 
$\Omega_\text{soft}(p,k)$ is given as the symmetric version of the dimensionally-regularized solution in eq.~(\ref{OmegaSoft}), where $(p^2/k^2)^\epsilon$ is replaced by 
$\left(\frac{(p^2)^2}{k^2(p-k)^2}\right)^{\epsilon/2}$. This guarantees that both soft limits are taken into account; moreover, $\Omega_\text{soft}(p,k)$ admits an $\epsilon$ expansion in terms of SVHPLs, and hence its two-dimensional limit reproduces exactly all non-vanishing terms in the $z\to 0$ and $z\to \infty$ limits of the two-dimensional solution $\Omega^{(\rm 2d)}(z,\bar{z})$ of (\ref{Omega2d}). As a consequence, one can isolate the two-dimensional limit of the hard wavefunction as follows: 
\begin{align}
\Omega_\text{hard}^{({\rm 2d})}(z,\bar{z})\, \equiv \,
\lim_{\epsilon \to 0} \Omega_{\text{hard}} 
= \Omega^{(\rm 2d)}(z,\bar{z}) - \Omega_{\text{soft}}^{({\rm 2d})}(z,\bar{z}).
\end{align}
Given that all the singularities in the amplitude 
arise from the soft component of the wavefunction, 
the full NLL amplitude (\ref{Amp}) to ${\cal O}(\eps)$ 
is given by 
\begin{align}
	\label{AmpComb}
	\hat{\cal M}^{(+),\rm NLL}_{ij\to ij}\left(\frac{s}{-t}\right) = -i\pi \left[ \int [{\rm D}k] \frac{p^2\, \Omega_{\rm soft}(p,k)}{k^2(p-k)^2} + \frac{1}{4\pi} \int \frac{d^2 z}{z\bar{z}} \Omega_{\rm hard}^{({\rm 2d})}(z,\bar{z}) \right] {\mathbf T}^2_{s-u} {\cal M}^{({\rm tree})}_{ij\to ij}\,.
\end{align}
The soft wavefunction is integrated in dimensional regularization 
in $d= 2-2\eps$ dimensions, while the hard wavefunction, which by 
construction vanishes in the soft limits, can be integrated in $d = 2$. 
We thus obtain the full NLL $2\to2$ amplitude, i.e. the leading tower 
of logarithms in the imaginary part of the amplitude. 
For the first few orders one has
\begin{align} \label{AmplRes} \nn
\hat{\cal M}^{(+,1,0)}_{ij\to ij} 
=&i\pi \frac{1}{2 \epsilon }{\mathbf T}^2_{s-u} {\cal M}^{({\rm tree})}_{ij\to ij}, 
\qquad \qquad
\hat{\cal M}^{(+,2,1)}_{ij\to ij} 
=i\pi C_2 \bigg[\frac{1}{8 \epsilon ^2}-\frac{ \zeta (2)}{8} \bigg]
{\mathbf T}^2_{s-u} {\cal M}^{({\rm tree})}_{ij\to ij},
\\
\hat{\cal M}^{(+,3,2)}_{ij\to ij} 
=&i\pi C_2^2 \bigg[\frac{1}{48 \epsilon ^3} -\frac{ \zeta (2)}{32 \epsilon} 
-\frac{29}{48} \zeta (3)\bigg]{\mathbf T}^2_{s-u} {\cal M}^{({\rm tree})}_{ij\to ij},
\\ \nn
\hat{\cal M}^{(+,4,3)}_{ij\to ij} 
=&i\pi C_2^2 \bigg[ \frac{C_2}{384 \epsilon ^4}-\frac{C_2 \zeta (2)}{192 \epsilon ^2} 
-\left(\frac{7C_2 }{288} +\frac{C_A }{192}  \right)
\frac{\zeta (3)}{\epsilon }-\frac{C_2 \zeta (4)}{48}
-\frac{C_A \zeta (4)}{128}\bigg]{\mathbf T}^2_{s-u} {\cal M}^{({\rm tree})}_{ij\to ij}.
\end{align}
The wavefunction is given up to $\ell=13$ loops in~\cite{Caron-Huot:2020grv}; higher orders can be obtained with the same algorithm. The amplitude coefficients $\hat{\cal M}^{(+,\ell,\ell-1)}$ have uniform transcendental weight, and are expressed in terms of single- and multiple-zeta values: the latter appear starting at 11 loops, with $\zeta_{5,3,3}$ being the first one to emerge. Such multiple zeta values originate from the hard part of the amplitude in \eqn{AmpComb}, and are thus only of the type that originates in SVHPLs. 

The singularities of $\hat{\cal M}^{(+,\ell,\ell-1)}$ are part of the soft contribution, and are thus resummed by \eqn{softAmplRes}. Combining this information with the factorization theorem for infrared singularities allows one to extract the corresponding soft anomalous dimension and sum it to all orders, as obtained in \cite{Caron-Huot:2017zfo}. In short, it is well known that infrared divergences in amplitudes factorize and exponentiate~\cite{Catani:1998bh,Sterman:2002qn,Aybat:2006wq,Aybat:2006mz,Gardi:2009qi,Gardi:2009zv,Becher:2009cu} according to   
\begin{equation}
\mathcal{M} = \mathbf{Z}\,\cdot \mathcal{H}, \qquad 
\mathbf{Z} =  \mathcal{P}\exp\left\{-\frac{1}{2}\int_0^{\mu^2}
\frac{d\lambda^2}{\lambda^2}\mathbf{\Gamma}\right\}\,,
\label{eq:IRfactor}
\end{equation}
where $\mathcal{H}$ is an infrared-renormalized hard amplitude, which is finite, and  $\mathbf{\Gamma}$ is the soft anomalous dimension. In the high-energy limit the 
latter takes the form~\cite{DelDuca:2011ae}
\begin{equation}\label{softADdef}
\mathbf{\Gamma} =\frac{\gamma_K}{2}\left[L\tts + i\pi\tsu\right] +  \Gamma_i + \Gamma_j+\mathbf{\Delta},
\end{equation}
where $\Gamma_i$ is defined in~(\ref{Zi}) and $\mathbf{\Delta}$ represents non-dipole corrections starting at three loops~\cite{Gardi:2009qi,Gardi:2009zv,Becher:2009cu,Almelid:2015jia}.
The dipole contribution is well known \cite{DelDuca:2011ae,DelDuca:2014cya}: the $L\tts$ term contributes starting at LL, while the $i\pi\tsu$ and $\Gamma_{i}$ terms start at NLL. Here we focus on $\mathbf{\Delta}$, which we expand as
\begin{equation}
\mathbf{\Delta} = \sum_{n=3}^\infty  \left(\frac{\al_s}{\pi}\right)^n\sum_{m=0}^{n-1}L^m
\mathbf{\Delta}^{(n,m)}\,.\label{eq:expansionDelta}
\end{equation}
The calculation of the tower $\hat{\cal M}^{(+,\ell,\ell-1)}$ 
allows us to determine the coefficients $\mathbf{\Delta}^{(-,\ell,\ell-1)}$ 
for all $\ell$. We find \cite{Caron-Huot:2017zfo}
\beq\label{GammaNLL3a}
\mathbf{\Delta}^{(-,\ell,\ell-1)} = i\pi \,G^{(\ell)} \,\Tsu, 
\quad {\rm with} \quad 
G^{(\ell)} \equiv \frac{1}{(\ell-1)!}\left[ \frac{(C_A-\Tt)}{2}\right]^{\ell-1} 
\left.\left( 1 - R(\epsilon) \frac{C_A}{C_A -\Tt} \right)^{-1}\right\vert_{\epsilon^{\ell-1}} ,
\end{equation}
where $R(\epsilon)$ is given in \eqn{Reps}, and the subscript $\vert_{\epsilon^{\ell-1}}$ indicates that one should extract the coefficient of $\epsilon^{\ell-1}$. The result in \eqn{GammaNLL3a} can be resummed to all orders: one finds 
\begin{equation} \label{GammaNLL3}
\mathbf{\Delta}_{\rm NLL}^{(-)} = i\pi \frac{\alpha_s}{\pi} 
\,G\left(\frac{\alpha_s}{\pi}L\right)\Tsu\,,
\end{equation}
where $G(x) = \sum_{\ell=1}^\infty x^{\ell-1} G^{(\ell)}$. It is interesting to note~\cite{Caron-Huot:2017zfo} that $G(x)$ is an entire function, thus it admits an infinite radius of convergence, which implies that one can compute the NLL soft anomalous dimension at any value of the effective coupling, including $\frac{\alpha_s}{\pi}L\gg1$. 

\begin{figure}[t]
\centering
\includegraphics[width=0.45\textwidth]{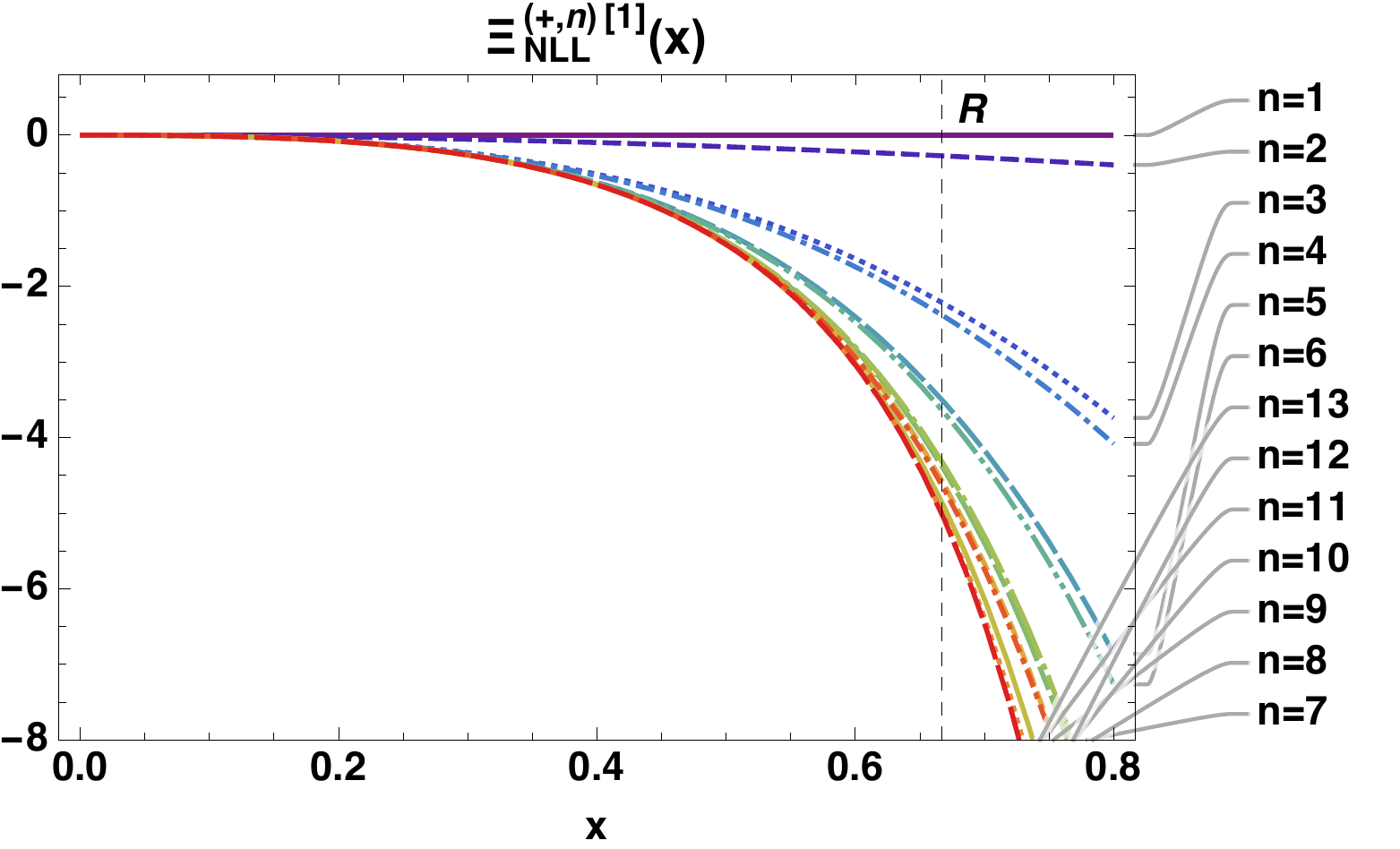}  
\includegraphics[width=0.45\textwidth]{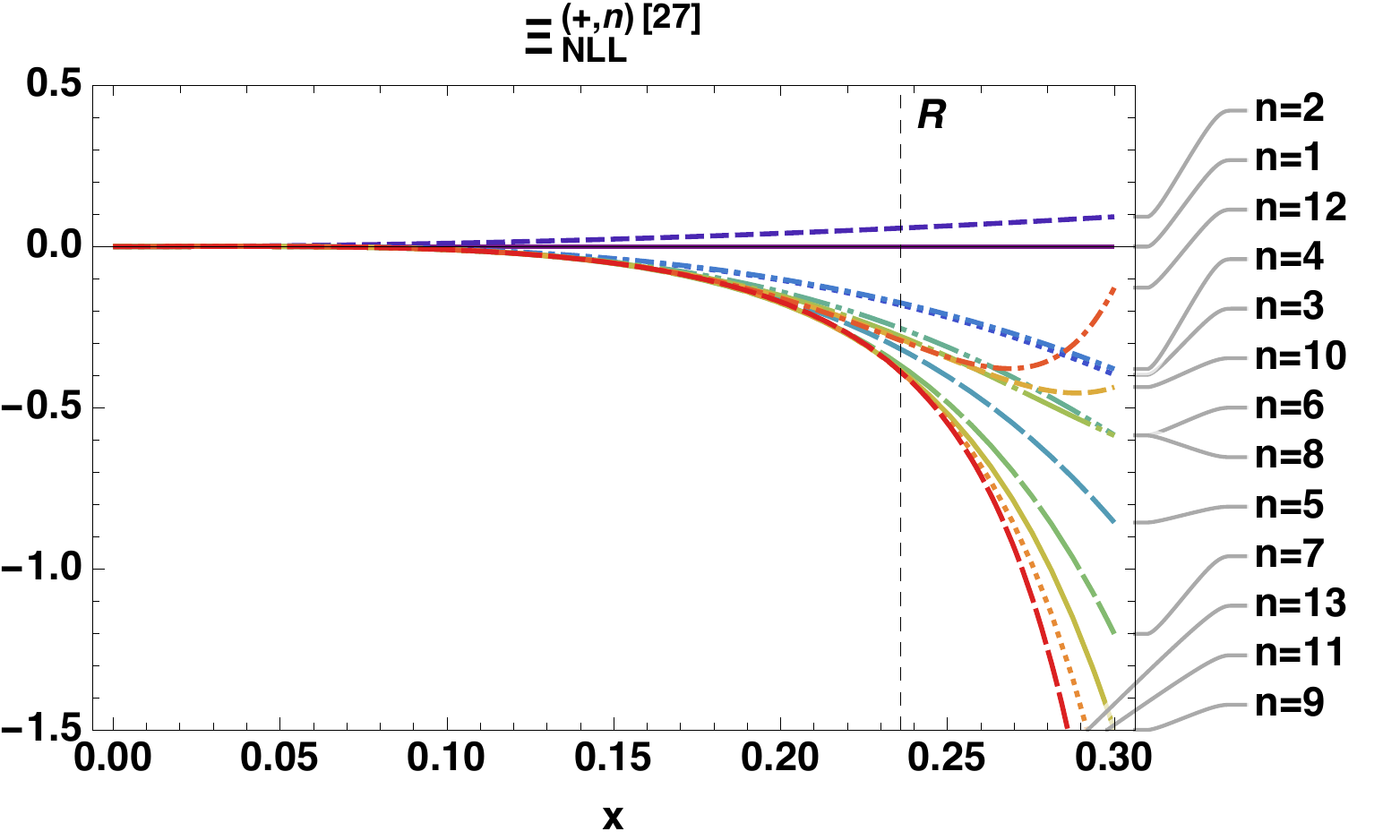}
\caption{Partial sums of the amplitude coefficients $\Xi_{\rm NLL}^{(+,\ell)}$, 
	up to 13th order, for the singlet (left plot) and the 27 colour representation (right plot). 
	The horizontal axis $x$ represents $\frac{\alpha_s}{\pi}L$. The dashed  
	vertical line represents the radius of convergence, $R$, determined by 
	the  pole closest to $x = 0$, using Pad\'{e} approximants.}
\label{Radius-Full}
\end{figure}

The finite, ${\cal O}(\epsilon^0)$ contribution to the NLL amplitude (\ref{AmplRes}) can be written as
\begin{equation}
	\hat{{\cal M}}^{(+)}_{\rm NLL}  = \frac{i \pi}{L} \, \Xi_{\rm NLL}^{(+)} \, \Tsu {\cal M}_{\rm tree}\,.
\end{equation}
It is not yet known how to resum $\Xi_{\rm NLL}^{(+)}$. In any case, it is clear that resummation would not involve $\Gamma$ functions only, because $\Xi_{\rm NLL}^{(+)}$ contains multiple zeta values. However, it is possible to study numerically the convergence of the series given by the coefficients $\Xi_{\rm NLL}^{(+,\ell,\ell-1)}$, for the relevant representations for $N_c=3$, the singlet (${\mathbf T}_{t}^2 \, {\cal M}^{[1]} = 0$) and the 27 representation (${\mathbf T}_{t}^2 \, {\cal M}^{[27]}  = 2(N_c+1)\, {\cal M}^{[27]}$). One finds:
\begin{align} 
\label{Xi1}  
\Xi_{\rm NLL}^{(+)[1]} &= -0.6169 \, x^2 - 6.536 \, x^3 
- 0.8371 \, x^4 - 8.483 \, x^5 - 1.529 \, x^6 - 12.67 \, x^7 \\ \nn
&\hspace*{-20pt}+\, 1.610 \, x^8- 20.62 \,  x^9 + 16.48\,  x^{10} 
- 35.98 \, x^{11} + 46.07 \, x^{12} 
- 74.04 \, x^{13} + {\mathcal{O}}(x^{14}), \\[0.1cm]
\label{Xi27} 
\Xi_{\rm NLL}^{(+)[27]} &= 1.028 \, x^2 - 18.16 \, x^3 
+ 2.184 \, x^4 - 196.0 \, x^5 + 372.3 \, x^6 - 2821 \, x^7  + 9382 \,  x^8 \\ \nn
&\hspace*{-20pt} -\, 46494 \, x^9  + 180397 \, x^{10}  - 797524 \, x^{11} 
+ 3.239 \times 10^{6} \,  x^{12} - 1.374 \times 10^{7} \, x^{13} + {\mathcal{O}}(x^{14}). 
\end{align}
By using Pad\'{e} approximants, it is possible to conclude~\cite{Caron-Huot:2020grv} that the series has a finite radius of convergence, as is illustrated in fig.~\ref{Radius-Full}.
We find asymptotic geometric progression with the powers of 
$-\frac12 C_2\frac{\alpha_s}{\pi}L=-\frac32 x$ and $({C_2}- \frac{3}{8}{C_1})  \frac{\alpha_s}{\pi}L=-\frac{17}{4}x$ for the singlet and the 27 representations, respectively. 
In both cases the series displays sign-oscillations, indicating that once resummed, it could be extrapolated beyond its radius of convergence.

\section{Three-Reggeon cut}
\label{sec:threeReggeonCut}

We turn now to considering the odd amplitude at NNLL accuracy, i.e., the tower of coefficients ${\mm}^{(-,n,n-2)}$. This part of the amplitude is particularly interesting from the point of view of Regge theory, because it involves a Regge cut in the real part of the amplitude, and it has quite different factorization properties compared to the two-Reggeon cut discussed so far, which contributes to the imaginary part of the amplitude. In this respect, an aspect of particular interest is the mixing between the single- and triple-Reggeon exchange, as discussed at the end of section \ref{sec:intro}, for which we refer the reader to \cite{upcoming}. In this talk we focus on the calculation of ${\mm}^{(-,n,n-2)}$, which follows from the expansion to NNLL accuracy of  
\beq \label{Regge-odd-Amplitude}
\frac{i}{2s} {\cal \hat M}^{(-)}_{ij\to ij} = 
\bra{\psi^{(-)}_j}e^{-\hat H L}\ket{\psi^{(-)}_i}
= D_i(t) D_j(t){\cal M}^{\rm tree}_{ij\to ij}
+ {\hat{{\cal M}}}^{(-),\,\rm MRS}_{ij\to ij}\,,
\eeq
where the second equality follows from \eqn{Schemes_MRS} and the definition of reduced amplitude in \eqn{eq:MhatDef}. As shown in \cite{Falcioni:2020lvv,upcoming}, this matrix element can be written to all orders as the sum of four contributions. In general, the NNLL contribution to the odd amplitude at $n$-loop is proportional to $L^{n-2}$, which is obtained from $n-2$ repeated actions of the Balitsky-JIMWLK Hamiltonian $H$. The diagonal transitions $H_{k\to k}$ in \eqn{Hamiltonian_schematic_form} are $\mathcal{O}(\al_s)$, while next-to-diagonal transitions $H_{k\to k\pm 2n}$ with $n=1$ are $\mathcal{O}(\al_s^2)$. Noting that $\ket{\psi_{i,n}}$ is $\mathcal{O}(g_s^{n-1})$, there are four different types of contributions $\mathcal{O}(\al_s^{n}L^{n-2})$:
\begin{align}\label{braket1331}
\begin{array}{cc}
\braket{\psi_{j,3}| \hat H_{3\to3}^k|\psi_{i,3}}, & 
\qquad\qquad 
\braket{\psi_{j,1}| \hat H_{3\to1}\, \hat H_{3\to3}^{k-2}\, 
	\hat H_{1\to3}|\psi_{i,1}}, \\[0.2cm]
\braket{\psi_{j,1}| \hat  H_{3\to1}\, \hat  H_{3\to3}^{k-1}|\psi_{i,3}}, & 
\qquad\qquad 
	\braket{\psi_{j,3}| \hat H_{3\to3}^{k-1} \, \hat H_{1\to3}|\psi_{i,1}},
\end{array}
\end{align}
where the two transitions in the second line of \eqn{braket1331} are related by target-projectile symmetry. These contributions can be expressed in a compact form by making the power-counting in the strong coupling manifest. To this end we introduce normalized Reggeon projectile states $\ket{i_n}$ (and similarly $\langle j|$ for the target) and a normalized reduced Hamiltonian $\tilde{H}$, defined as 
\begin{align}\label{eq:newNormalisations}
(r_\Gamma\al_s)^{(n-1)/2} \,\ket{i_n} \, \equiv\ket{\psi_{i,n}}, 
\quad \qquad 
\left(\frac{\al_s}{\pi} r_\Gamma\right)^{1+n} \,\pi^n
\, {\tilde H}_{k\to k+2n}  \,  \equiv \, 
{\hat{H}}_{k\to k+2n}\,,
\end{align}
where we introduce rescalings by the constant $r_\Gamma$ 
defined in \eqn{al_g}, and in turn we expand $\ket{i_{n}} 
=\ket{i_{n}}^{\text{LO}} +\sum_{k=1}^{\infty}
\left(\frac{\alpha_s}{\pi}\right)^k 
\ket{i_{n}}^{\text{N}^k\text{LO}}$. With this notation, 
the four terms in \eqn{braket1331} give rise to the NNLL 
odd amplitude as follows:
\begin{align}
\begin{split}
	\label{eq:mhatNNLL}
	\frac{i}{2s}\hat{\mm}^{(-),\text{NNLL}}_{ij\to ij} =& \,
	\pi^2 \Bigg[\sum_{\ell=0}^{\infty}\frac{(-L)^\ell}{\ell!}
	\left(\frac{\al_s}{\pi}r_\Gamma \right)^{\ell+2}
	\braket{j_{3}|\tilde{H}_{3\to3}^\ell|i_{3}}  \\ 
	&\hspace{-1.6cm}+\, \sum_{\ell=1}^{\infty}
	\frac{(-L)^{\ell}}{\ell!}\left(\frac{\al_s}{\pi}
	r_\Gamma\right)^{\ell+2} \left[
	\braket{j_{1}|\tilde{H}_{3\to1}\tilde{H}_{3\to3}^{\ell-1}|i_{3}}
	+\braket{j_{3}|\tilde{H}_{3\to3}^{\ell-1}\tilde{H}_{1\to3}|i_{1}}\right] \\ 
	&\hspace{-1.6cm} +\, \sum_{\ell=2}^{\infty}
	\frac{(-L)^{\ell}}{\ell!}\left(\frac{\al_s}{\pi} r_\Gamma\right)^{\ell+2}
	\braket{j_{1}|\tilde{H}_{3\to1}\tilde{H}_{3\to3}^{\ell-2}\tilde{H}_{1\to3}|i_{1}}
	\Bigg]^{\text{LO}}\!\!\!\!
	+\left(\frac{\alpha_s}{\pi}\right)^2
	\braket{j_{1}|i_{1}}^{\text{NNLO}}\!\!.
\end{split}    
\end{align}
Notice that all but the last term are described within the leading order (LO) formalism, where the matter content of the theory is irrelevant. The matrix elements are given in terms of LO projectile/target states with either one or three Reggeons, which in turn are evolved by means of the LO Balitsky-JIMWLK Hamiltonian. The two-loop contribution $\braket{j_{1}|i_{1}}^{\text{NNLO}}$, instead, needs to be extracted from  two-loop amplitudes, see ref.~\cite{Caron-Huot:2017fxr}, but it does not enter the calculation of reduced amplitudes to higher orders. Apart from $\braket{j_{1}|i_{1}}^{\text{NNLO}}$, eq.~(\ref{eq:mhatNNLL}) is therefore universal, in that it applies in any gauge theory, fully governed by the LO Balitsky-JIMWLK evolution of infinite Wilson lines in eq.~(\ref{rapidity_evolution}).

Eq.~\eqref{eq:mhatNNLL} has characteristic analytic properties. First of all, it is interesting to notice that the NNLL amplitude is proportional, to all orders, to a factor of $\pi^2$, if we exclude the two-loop single-Reggeon contribution. This is a characteristic feature of the three-Reggeon cut, playing a role similar to the universal factor of $i \pi$ appearing in case of the two-Reggeon cut in \eqns{Amp}{AmplRes}. Furthermore, as in case of the two-Reggeon cut \eqn{AmplRes}, based on form of the LO Balitsky-JIMWLK Hamiltonian, we expect the coefficients ${\mm}^{(-,n,n-2)}$ to have maximal, uniform transcendental weight, when the dimensional regularization parameter $\epsilon$ is counted as having weight $-1$.

\Eqn{eq:mhatNNLL} can be calculated iteratively. Compared to the computation of the two-Reggeon cut discussed in section \ref{sec:twoReggeonCut}, the integrals involved in the calculation of \eqn{eq:mhatNNLL} are of similar complexity; however, the colour structure is more involved, and it
is non trivial to express it in terms of colour operators acting on the tree level. For this purpose, specific techniques and new colour identities need to be developed. This task has been carried out in 
\cite{Falcioni:2020lvv,upcoming} where the odd amplitude has been determined explicitly up to four loop. For a detailed discussion of these techniques we refer the reader to these works. In the following we focus on presenting the result for the odd amplitude up to four loops. 

The two and three-loops amplitude coefficients were already obtained in \cite{Caron-Huot:2017fxr}. Using the notation of \eqn{eq:mhatNNLL}, at the two-loop amplitude has two contributions:
\begin{equation}
\label{reduced_ampl_two_loop}
\frac{i}{2s}\hat{\mathcal{M}}^{(-,2,0)}_{ij\to ij} = \langle j_1|i_1\rangle^{\text{NNLO}} + \pi^2r_\Gamma^2\,\langle j_3|i_3\rangle.
\end{equation}
The single-Reggeon transition, comparing with \eqn{Regge-odd-Amplitude}, reads
\begin{equation}
	\label{eq:resJ1I1NNLO}
	\langle j_1 | i_1\rangle^{\text{NNLO}} = \left[D_i^{(2)}+D_j^{(2)}+D_i^{(1)}D_j^{(1)}\right]\langle j_1 | i_1\rangle.
\end{equation}
The three-reggeon transition gives
\begin{equation}
\label{eq:resJ3I3}
\langle j_3 | i_3\rangle =-\frac{1}{8}\left(\frac{1}{\epsilon^2}-6\epsilon\,\hat{\zeta}_3
+{\cal{O}}\left(\epsilon^3\right)\right)\,
\left[\left(\tsu\right)^2-\frac{C_A^2}{12}\right]\langle j_1 | i_1\rangle,
\end{equation}
where the tree-level factor $\langle j_1 | i_1\rangle$ reads, according to the normalization of the states in \eqn{eq:newNormalisations}, $\langle j_1 | i_1\rangle = \frac{i}{2s}\hat{\mathcal{M}}^{\text{tree}}_{ij\to ij}$. Furthermore, $\hat{\zeta}_3 = \zeta_3 + \frac{3}{2}\epsilon\zeta_4 - \frac{5}{2}\epsilon^3\zeta_6+ {\cal{O}}\left(\epsilon^5\right)$. Summing the two contributions, 
one has 
\begin{equation}
\mExpM{2}{0} =\left[D_i^{(2)}+D_j^{(2)}+D_i^{(1)}D_j^{(1)}+\pi^2 r_\Gamma^2\,S^{(2)}(\epsilon)\left((\tsu)^2-\frac{1}{12}C_A^2\right)\right]\mTree\label{eq:M20},
\end{equation}
where 
\begin{equation}\label{eq:R2Res}
	S^{(2)}(\epsilon)= -\frac{1}{8\eps^2} +\frac{3}{4}\eps\zeta_3 + \frac{9}{8}\eps^2\zeta_4 +{\cal O}(\eps^3).
\end{equation}

At three loops the action of the Hamiltonian kicks-in, and one has to take into account three MRS contributions:
\begin{equation}
	\label{eq:expaMhatnnll3}
	\frac{i}{2s}\hat{\mathcal{M}}^{(-,3,1)}_{ij\to ij}= -\pi^2 r_\Gamma^3\Big[\langle j_3| \tilde{H}_{3\to3} | i_3\rangle\,+\,\left(\langle j_1 |\tilde{H}_{3\to1}|i_3\rangle + \langle j_3 |\tilde{H}_{1\to3}|i_1\rangle\right)\Big].
\end{equation}
The various matrix elements can be evaluated to give 
\begin{align}
\begin{split}
\label{eq:resJ3H33I3}
	\langle j_3 | \tilde{H}_{3\to3} | i_3\rangle &=-24\,S_C^{(3)}(\epsilon)\,\left(\frac{d_{AR_i}}{N_{R_i}C_i}+\frac{d_{AR_j}}{N_{R_j}C_j}\right)\langle j_1 | i_1\rangle\\
	&+\Big[S_A^{(3)}(\epsilon)\tsu[\tsu,\tts]+S_B^{(3)}(\epsilon)[\tsu,\tts]\tsu+S_C^{(3)}(\epsilon)C_A^3\Big]\,\langle j_1| i_1\rangle,
\end{split}    
\end{align}
and 
\beq
\label{eq:resJ1H31I3}
	\langle j_3 | \tilde{H}_{1\to3} | i_3\rangle + \langle j_1 |\tilde{H}_{3\to1}|i_3\rangle =24\,S_C^{(3)}(\epsilon)\,\left(\frac{d_{AR_i}}{N_{R_i}C_i}+\frac{d_{AR_j}}{N_{R_j}C_j}\right)\,\langle j_1|i_1\rangle\,,
\eeq
where $d_{AR_i}$ and $d_{AR_j}$ are quartic Casimir associated to the projectile and target:
\begin{equation}
	\label{eq:dARi}
	d_{AR_i}=\frac{1}{4!}\sum_{\sigma\in\mathcal{S}_4}\mathrm{Tr}\left[F^{\sigma(a)}F^{\sigma(b)}F^{\sigma(c)}F^{\sigma(d)}\right]\mathrm{Tr}\left[\mathbf{T}^a_i\mathbf{T}^{b}_i\mathbf{T}^{c}_i\mathbf{T}^{d}_i\right],
\end{equation}
and the functions $S_{i}^{(3)}(\eps)$ arise from the evaluation of loop integrals in $2-2\eps$ dimensions: 
\begin{subequations}
\label{RABC}
\begin{align}
\label{eq:defRA3}
S_A^{(3)}(\epsilon) & =
\frac{1}{48\epsilon^3}+\frac{37\,\hat{\zeta}_3}{24}\,+\,{\cal{O}}\left(\epsilon^2\right)\,,\\
\label{eq:defRB3}
S_B^{(3)}(\epsilon) & =
\frac{1}{24\epsilon^3} +\frac{\hat{\zeta}_3}{12} \,+\,{\cal{O}} \left(\epsilon^2\right)\,,\\
\label{eq:defRC3}
S_C^{(3)}(\epsilon) & =
-\frac{1}{432}\left(\frac{1}{2\epsilon^3}-35\hat{\zeta}_3 + {\cal{O}}\left(\epsilon^2\right)\right)\,.
\end{align}
\end{subequations}
It is interesting to note that the sum of the transitions $1 \to 3$ and $3 \to 1$ does not involve a matrix in colour space, but is simply proportional to $\braket{j_1|i_1}$. This is due to the fact that colour is carried by a single Reggeon on either the target or projectile sides. This property holds for all terms in the second and third lines of~(\ref{eq:mhatNNLL}) at any order. Another intriguing feature \cite{Falcioni:2020lvv,upcoming} is that the quartic Casimir contributions of \eqn{eq:resJ1H31I3}, exactly cancel an identical contribution in the $3\to 3$ transition. One then finds
\begin{equation}
	\mExpM{3}{1} =-\pi^2r_\Gamma^3\left[S_A^{(3)}(\epsilon)\tsu[\tsu,\tts]+S_B^{(3)}(\epsilon)[\tsu,\tts]\tsu+S_C^{(3)}(\epsilon)C_A^3\right]\mTree\label{eq:M31}.
\end{equation}

\begin{figure}
\begin{center}
\includegraphics[width=0.10\textwidth]{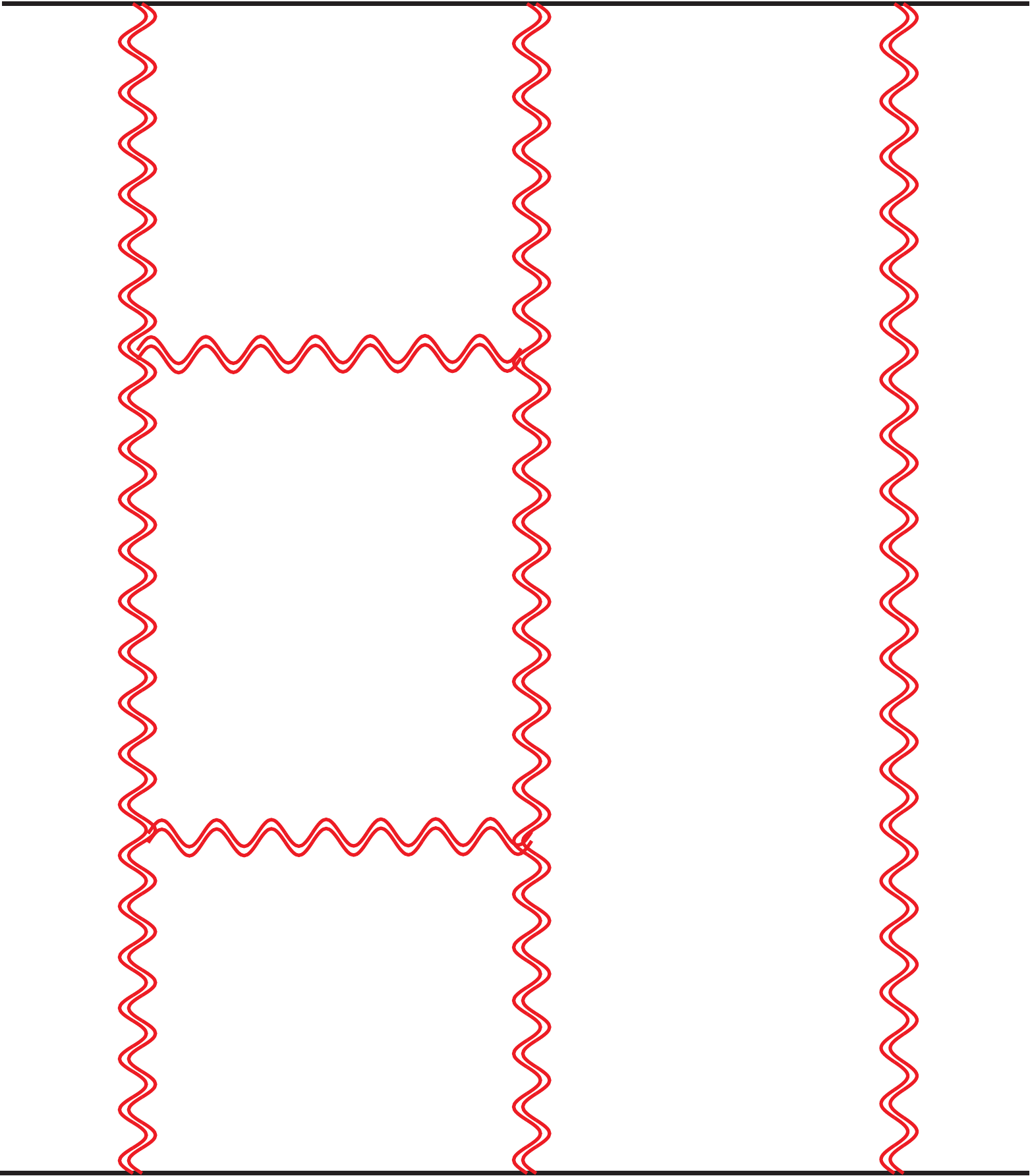}
\hspace{15pt}
\includegraphics[width=0.10\textwidth]{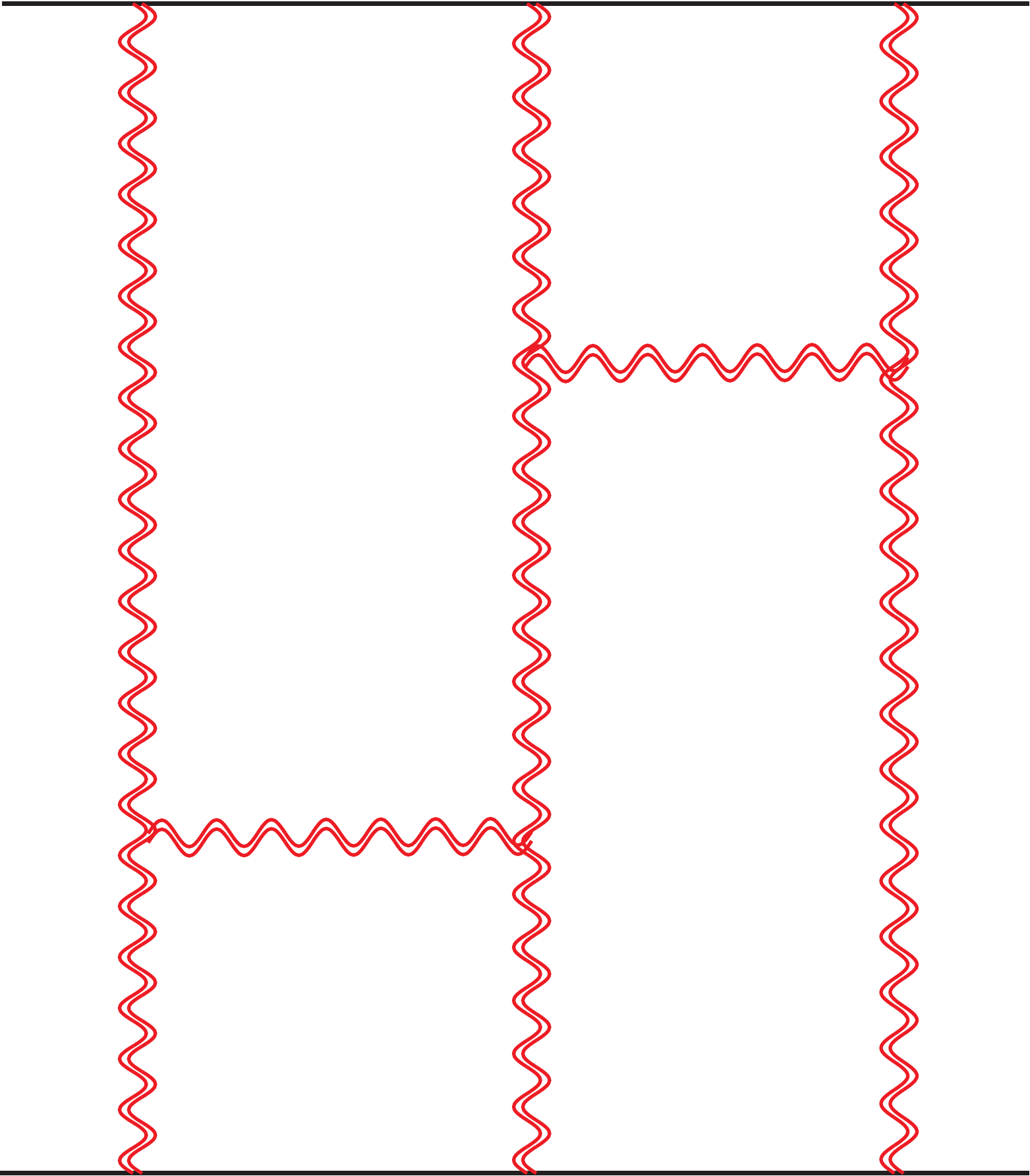}
\hspace{15pt}
\includegraphics[width=0.10\textwidth]{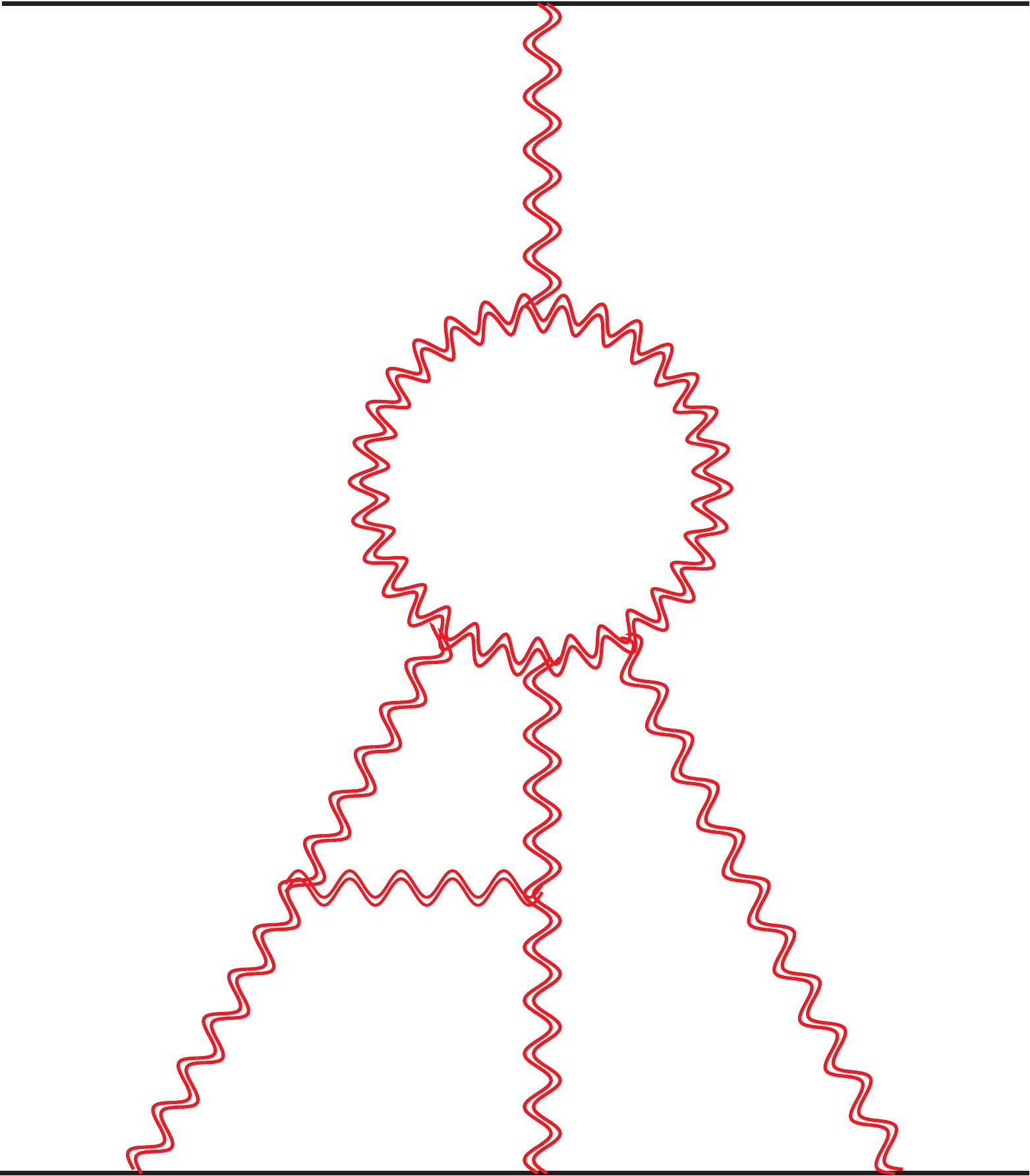}
\hspace{15pt}
\includegraphics[width=0.10\textwidth]{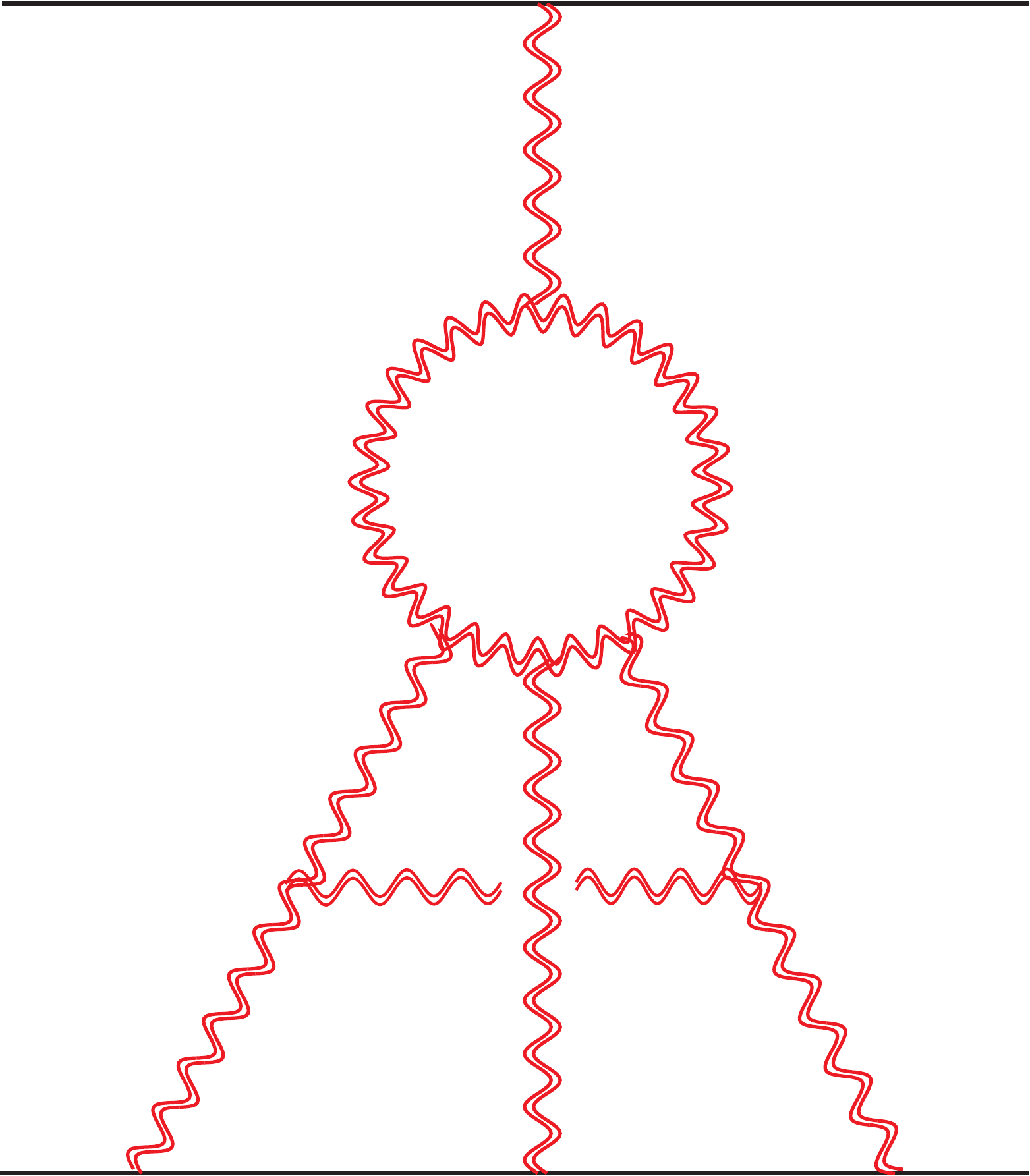}
\hspace{15pt}
\includegraphics[width=0.10\textwidth]{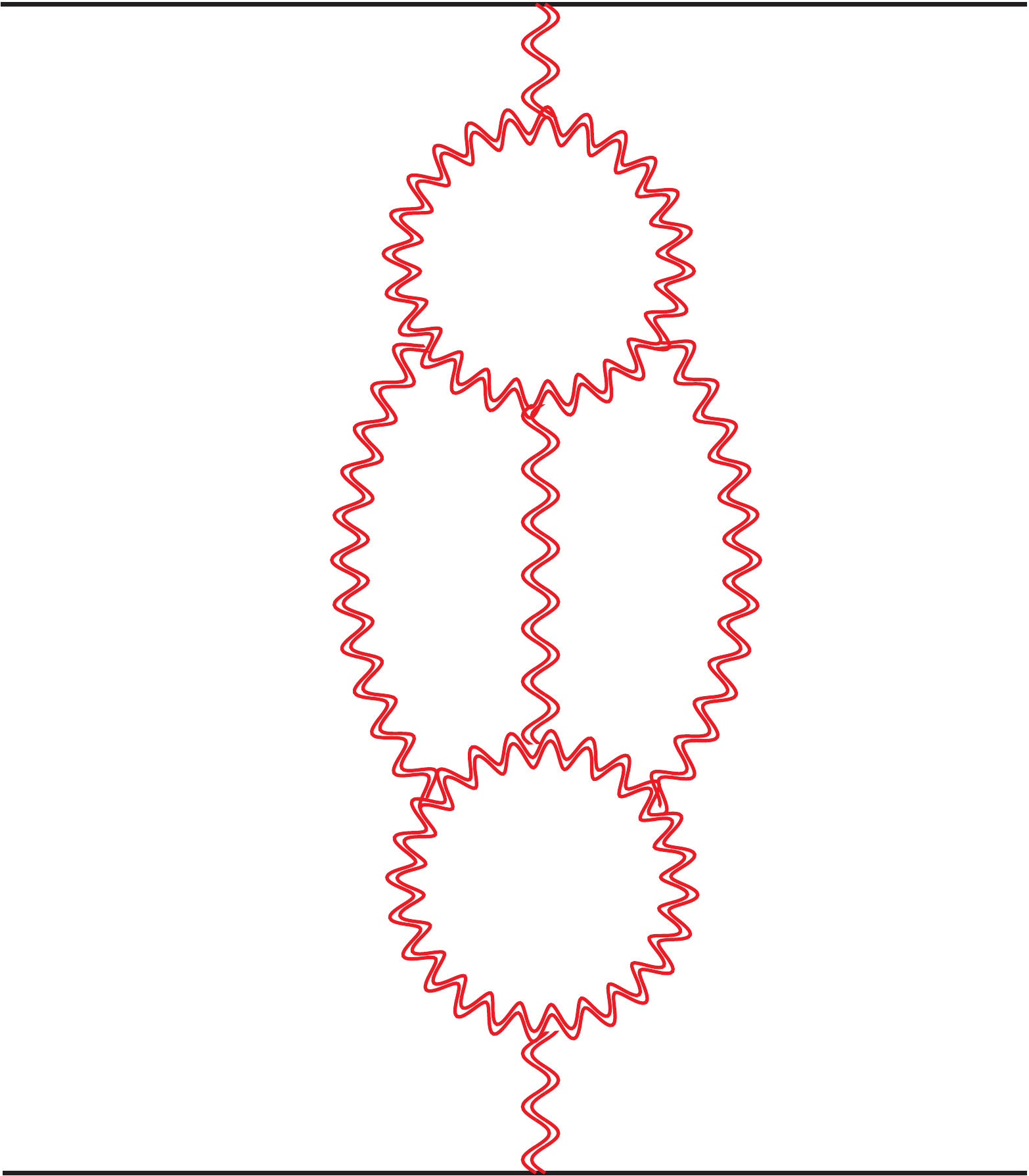}
\caption{Representation of transitions involving Multi-Reggeon States (MRS),
which contribute to the odd-signature amplitude at four loops.}
\label{fig:4loops}
\end{center}
\end{figure}
Next, we consider the four-loop amplitude, determined for the first time in \cite{Falcioni:2020lvv}. In this case we need to take into account two consecutive application of the Balitsky-JIMWLK Hamiltonian, which gives rise for the first time to all four MRS terms in \eqn{eq:resJ3I3}. One has 
\begin{align}
\begin{split}
	\label{eq:expaMhatnnll4}
	\frac{i}{2s}\hat{\mathcal{M}}^{(-,4,2)}_{ij\to ij} &= \frac{\pi^2r_\Gamma^4}{2}\;\Big[\langle j_3|\tilde{H}_{3\to3}^2|i_3\rangle + \left(\langle j_1| \tilde{H}_{3\to1}\tilde{H}_{3\to3}|i_3\rangle + \langle j_3 |\tilde{H}_{3\to3}\tilde{H}_{1\to3}|i_1\rangle \right) \\
	&\hspace{3.0cm}+ \langle j_1|\tilde{H}_{3\to1}\tilde{H}_{1\to3}|i_1\rangle \Big],
\end{split}
\end{align}
The first term, i.e. the double insertion of $\tilde{H}_{3\to3}$ in the three-Reggeon ladder, gives the most complicated contribution, which is schematically represented by the two diagrams on the left in fig. \ref{fig:4loops}. The first diagrams involve BFKL evolution applied twice to the same pair of Reggeons, while in the second diagram evolution is applied at each stage to a different pair. One obtains
\begin{align}
	\label{eq:reshat33hat33}
	\begin{split}
		\langle j_3|\tilde{H}_{3\to3}^2| i_3\rangle &=\Bigg[\frac{\mathbf{C}^{(4,-4)}}{\epsilon^4}+\frac{\hat{\zeta}_3}{\epsilon}\,\mathbf{C}^{(4,-1)}+{\cal{O}}\left(\epsilon\right)\Bigg]\,\langle j_1|i_1\rangle,
	\end{split}
\end{align}
where we have defined the following operators in colour space
\begin{align}
\begin{split}
\label{eq:resC44}
\mathbf{C}^{(4,-4)} =&\frac{1}{432}\left[\frac{d_{AA}}{N_A}
-3C_A\left(\frac{d_{AR_i}}{N_{R_i}C_i}+\frac{d_{AR_j}}{N_{R_j}C_j}\right)
+\frac{C_A^4}{12}\right]-\frac{1}{192}[\tsu,\tts]\tts\tsu \\
&+\frac{1}{96}\big[\tsu,[\tsu,\tts]\big]\tts
+\frac{7}{576}\tts[\big(\tsu\big)^2,\tts]
-\frac{5}{192}\tsu[\tsu,\tts]\tts,
\end{split}  \\  
\begin{split}
\label{eq:resC41}
\mathbf{C}^{(4,-1)} =&-\frac{101}{216}\left[\frac{d_{AA}}{N_A}
-\frac{312}{101}C_A\left(\frac{d_{AR_i}}{N_{R_i}C_i}
+\frac{d_{AR_j}}{N_{R_j}C_j}\right)+\frac{211C_A^4}{2424}\right]
+\frac{101}{96}[\tsu,\tts]\tts\tsu \\
&+\frac{49}{48}\Big[\tsu,\big[\tsu,\tts\big]\Big]\tts
-\frac{47}{288}\tts\,[\big(\tsu\big)^2,\tts]
-\frac{49}{48}\tsu[\tsu,\tts]\tts.
\end{split}
\end{align}
The terms in square brackets in eqs.~(\ref{eq:resC44}) and (\ref{eq:resC41}) are leading in the planar limit, while all other terms, involving commutators of $\tsu$ and $\tts$, are suppressed in $N_c$, and result in mixing of the octet exchange with other colour. 

The third and fourth diagrams in fig. \ref{fig:4loops} correspond to the $3 \to 1$ and $1\to 3$ transitions given by the second and third terms in \eqn{eq:expaMhatnnll4}. The last diagram correspond to the $1\to 1$ transition mediated by three Reggeons, i.e. the last term in \eqn{eq:expaMhatnnll4}. These contributions are all proportional to the unit matrix in colour space. For the $3 \to 1$ and $1\to 3$ transitions we obtain \cite{Falcioni:2020lvv,upcoming}
\begin{align}
\label{eq:resJ1H31H33I3simb}        
\begin{split}
\langle j_1 |\tilde{H}_{3\to1}\tilde{H}_{3\to3}|i_3\rangle &+ \langle j_3 |\tilde{H}_{3\to3}\tilde{H}_{1\to3}|i_1\rangle = \\ &=\frac{C_A}{144}\left(\frac{d_{AR_i}}{N_{R_i}C_i}+\frac{d_{AR_j}}{N_{R_j}C_j}\right)\left(\frac{1}{\epsilon^4}-\frac{208\hat{\zeta}_3}{\epsilon}+{\cal{O}}\left(\epsilon\right)\right)\langle j_1|i_1\rangle\,,
\end{split}
\end{align}
and for the $1 \to 1$ transition 
\begin{align}
\begin{split}
	\label{eq:resJ1H31H13I1}
	\langle j_1| \tilde{H}_{3\to1}\tilde{H}_{1\to3}|i_1\rangle&=\Bigg[-\frac{1}{432\epsilon^4}\left(\frac{d_{AA}}{N_A}+\frac{C_A^4}{12}\right)+\frac{55\,\hat{\zeta}_3}{108\epsilon}\left(\frac{d_{AA}}{N_A}+\frac{101C_A^4}{1320}\right)\Bigg]\langle j_1|i_1\rangle.
\end{split}    
\end{align}
Summing the contributions in 
eqs.~(\ref{eq:reshat33hat33}),~(\ref{eq:resJ1H31H33I3simb}) 
and~(\ref{eq:resJ1H31H13I1}) we get   
\begin{equation}
	\label{eq:M42}
	\mExpM{4}{2}=\frac{r^4_\Gamma\pi^2}{2}\Bigg[\frac{1}{\epsilon^4} \mathbf{K}^{(4)} +
	\left( \frac{1}{\epsilon} \zeta_3 + \frac{3}{2}\zeta_4 \right)
	\mathbf{K}^{(1)}+ {\cal O}(\epsilon) \Bigg]\mTree,
\end{equation}
where the colour structures are defined by
\begin{subequations}
\begin{align}
\begin{split}
\label{eq:K4}
\mathbf{K}^{(4)}&=\frac{1}{96}\big[\tsu,[\tsu,\tts]\big]\tts
+\frac{7}{576}\tts\big[(\tsu)^2,\tts] \\
&-\frac{1}{192}[\tsu,\tts]\tts\tsu
-\frac{5}{192}\tsu[\tsu,\tts]\tts,
\end{split} \\
\begin{split}
\label{eq:K1pole}
\mathbf{K}^{(1)}&=\frac{49}{48}\big[\tsu,[\tsu,\tts]\big]\tts
-\frac{47}{288}\tts\big[(\tsu)^2,\tts]\\
&+\frac{101}{96}[\tsu,\tts]\tts\tsu-\frac{49}{48}\tsu[\tsu,\tts]\tts 
+ \frac{1}{24}\left(\frac{d_{AA}}{N_A}-\frac{C_A^4}{24}\right)\,. 
\end{split}
\end{align}
\end{subequations}
Eq.~(\ref{eq:K1pole}) depends on the quartic Casimir $d_{AA}$, irrespective of the representation associated to the target and to the projectile. The expressions in eq.~(\ref{eq:K4}) and (\ref{eq:K1pole}) imply that the reduced amplitude $\mExpM{4}{2}$ is non-planar, which is expected on general grounds, \cite{upcoming}, based on its direct connection to the Regge cut. This is because eq.~(\ref{eq:K4}) is written in terms of the commutator $[\tts,\tsu]$, which is subleading in $N_c$, and the only non-commutator term is also subleading, given that 
\begin{equation}\label{eq:dAAminusCa4}
	\frac{d_{AA}}{N_A}-\frac{C_A^4}{24} = 0\cdot N_c^4+\frac{3}{2}N_c^2.
\end{equation}

We note that, as at three loops, all terms arising from $3\to 1$ and $1\to 3$ transitions, which are proportional to the quartic Casimir invariants $d_{AR_i}$ and $d_{AR_j}$, cancel in the sum with the $3\to3$ amplitude, to all orders in $\epsilon$. These contributions are separately leading in the planar limit, thus their cancellation is essential for the complete $\hat{\mathcal{M}}^{(-,4,2)}$ to be non-planar. Based on this observation, we conjecture that transition amplitudes connecting three-Reggeon states with a single Reggeon cancel from the reduced amplitude $\hat{\mathcal{M}}^{(-),\text{NNLL}}$ to all perturbative orders.

With the odd amplitude coefficients ${\mm}^{(-,n,n-2)}$ evaluated up to four-loop, several aspects concerning the analytic structure of two-parton amplitudes can be investigated. We refer to \cite{upcoming} for a throughout analysis. Here we limit ourselves to one application, namely, the extraction of the soft anomalous dimension, (see refs. \cite{upcoming,Maher:2021nlo}) which can be obtained by combining the knowledge of the amplitude in the high-energy limit with the infrared factorization theorem, as discussed for the even amplitude at NLL accuracy. In this case, the calculation of the tower ${\mm}^{(-,\ell,\ell-2)}$ gives access to $\mathbf{\Delta}^{(+,\ell,\ell-2)}$. Here we have determined ${\mm}^{(-,\ell,\ell-2)}$ up to $\ell = 4$, therefore we obtain \cite{Caron-Huot:2017fxr,Falcioni:2020lvv,upcoming}:
\begin{align}\label{GammaNNLL3}
\mathbf{\Delta}^{(+,3,1)} & = 0, \\
\label{GammaNNLL4}
\mathbf{\Delta}^{(+,4,2)} & =
\zeta_2 \zeta_3 \bigg(\frac{d_{AA}}{N_A} -\frac{C_A^4}{24}-\frac{1}{4}\tts[(\tsu)^2,\tts]+\frac{3}{4}[\tsu,\tts]\tts\tsu\bigg).
\end{align}
Thanks to the calculation of the four-loop amplitude, we have obtained the first non-trivial real contribution to the tower $\mathbf{\Delta}^{(+,\ell,\ell-2)}$. For completeness, we recall that also the imaginary contribution $\mathbf{\Delta}^{(-,3,1)}$ has been determined from the three-loop exact calculation in general kinematic, \cite{Almelid:2015jia}, and reads $\mathbf{\Delta}^{(3,1)} = i\pi\frac{\zeta_3}{4}[\tts,[\tts,\tsu]]$ \cite{Caron-Huot:2017fxr}. We refer to \cite{upcoming} for a complete list of known coefficients $\mathbf{\Delta}^{(\pm,\ell,m)}$ in the high-energy limit. 
 
With the result for the soft anomalous dimension, it is easy to invert \eqn{eq:IRfactor} and obtain a prediction for the finite reminder of the amplitude at four loop. To this end, given the reduced amplitude in \eqn{eq:M42}, one needs first to restore the single Reggeon contribution which had been removed according to \eqn{eq:MhatDef}. The finite reminder then reads  
\begin{equation}\label{eq:IRinvert}
\mathcal{H} = Z_iZ_j\mathbf{Z}^{-1}e^{\al_g\tts L}\hat{\mathcal{M}}.
\end{equation}
Upon expansion the four-loop contribution at ${\cal O}(\eps^0)$ we get 
\begin{align}\label{eq:hardFunctionTheoryDependence}
\mathcal{H}^{(-,4,2)}=&\bigg\{\frac{C_A^2}{2}\left(\ag{2,0}\right)^2
+\frac{3}{16}\zeta_4\zeta_2 \, C_{\bf \Delta}^{(+,4,2)}
+\mathcal{O}(\eps)\bigg\}\mTree,
\end{align}
where $\ag{2,0} = C_A \left(\frac{101}{108} - \frac{\zeta_3}{8}\right) - \frac{7}{27}T_R n_f$ is the $\mathcal{O}(\eps^0)$ term of the two-loop Regge trajectory, and 
\beq
C_{\bf \Delta}^{(+,4,2)} \equiv \frac{d_{AA}}{N_A}-\frac{C_A^4}{24} 
+ \frac{1}{4}\tts[\tts,(\tsu)^2] +\frac{3}{4}[\tsu,\tts]\tts\tsu .
\eeq
Eq.~(\ref{eq:hardFunctionTheoryDependence}) conveniently displays the theory dependence of $\mathcal{H}^{(-,4,2)}$, which is in fact restricted to $\ag{2,0}$. We can thus obtain explicit result in QCD:
\begin{align}
\begin{split} \label{eq:QCDHard}
\mathcal{H}^{(-,4,2)}_{\text{QCD}}=&\,
\bigg\{ C_A^2 T_R^2n_f^2\frac{49}{1458}+C_A^3 T_Rn_f 
\left(\frac{7\zeta_3}{216}-\frac{707}{2916}\right)
+C_A^4 \left(\frac{\zeta_3^2}{128}
-\frac{101\zeta_3}{864}+\frac{10201}{23328}\right) \\
&\hspace{4.0cm}
+\frac{3}{16}\zeta_4\zeta_2 \, C_{\bf \Delta}^{(+,4,2)}
+\mathcal{O}(\eps)\bigg\}\mTree,
\end{split}
\end{align}
which is a new result. Furthermore, according to the principle of 
maximum trascendentality, in $\mathcal{N}=4$ SYM
one has:
\begin{align}
\begin{split}\label{eq:hardFuncRes} 
\mathcal{H}^{(-,4,2)}_{\text{SYM}}=&\,
\bigg\{\frac{C_A^4}{128}\zeta_3^2
+\frac{3}{16}\zeta_4\zeta_2\, C_{\bf \Delta}^{(+,4,2)}
+\mathcal{O}(\eps)\bigg\}\mTree.
\end{split}
\end{align}
While for $\mathcal{N}=4$ SYM the planar limit, 
$N_c\to\infty$, is already known 
\cite{Anastasiou:2003kj,Bern:2005iz,Drummond:2007aua}, 
the non-planar correction, i.e. the second term in 
\eqn{eq:hardFuncRes} is new.

\section{Conclusion}

In this talk we have summarized recent progress concerning the calculation of amplitudes in the high-energy limit. Within the shockwave formalism, scattering amplitudes can be calculated as expectation values of Wilson lines associated to the external partons \cite{Balitsky:1995ub,Balitsky:2001mr}. In this framework one can identify an effective degree of freedom, identified as a ``Reggeon'' \cite{Caron-Huot:2013fea}, which plays a central role in determining the factorization properties of the amplitude in the high-energy limit. Upon expansion in perturbation theory, the amplitude can be described in terms of transition amplitudes between states labelled by the number Reggeons, which needs to be evolved to equal rapidity by means of the BFKL equation and its generalization, the Balitsky-JIMWLK equation \cite{Caron-Huot:2013fea,Caron-Huot:2017fxr}. While it is not known how to solve these equations exactly, it is possible to develop iterative solutions, which allows one to calculate the amplitude to high order in perturbation theory. 

In this context, we have discussed recent calculations of $2\to 2$ scattering amplitudes in the high-energy limit. Refs. \cite{Caron-Huot:2017zfo,Caron-Huot:2020grv} focus on the imaginary part of the amplitude at NLL accuracy in the high-energy logarithm. Within the shockwave formalism this can be expressed as a class of ladder graphs involving two Reggons with any number of rungs, which arise from the iterative solution of the BFKL equation. We have shown that the singular part \cite{Caron-Huot:2017zfo} can be associated to a kinematic configuration in which one of the two Reggeons become soft. Within this approximation the evolution equation simplifies, such that the singular part of the amplitude can be calculated to all order and resummed in a closed form involving only gamma functions. We have then discussed the calculation of the finite contribution, for which a general algorithm has been set up, based on an iterative solution of the BFKL equation in two transverse dimensions, where the calculation is greatly simplified by the fact that the two-dimensional two-Reggon wavefunction consists of single-valued HPLs \cite{Caron-Huot:2020grv}. Explicit results have been presented up to 13 loops order. Finite corrections to the amplitude have a more complicated pattern, involving multiple zeta values, and hence they cannot be resummed in terms of gamma functions.

We moved then to discuss the calculation of the real part of the two-parton scattering amplitude at NNLL accuracy, recently investigated in \cite{Falcioni:2020lvv,upcoming}. We have shown that the amplitude can be expressed to all order as an iterated solution of the Balitsky-JIMWLK equation, with a simple diagrammatic interpretation in terms of ladder graphs involving three Reggons, with any number of rungs. Remarkably, the calculation of the entire tower requires only the LO Balitsky-JIMWLK formalism, while the complete odd amplitude at NNLL requires only two  additional inputs, given by the single-Reggeon impact factors at two loops and Regge trajectory at three loops, which can be determined by matching with the full amplitude. In \cite{Falcioni:2020lvv,upcoming} the signature-odd amplitude has been calculated for the first time to four loops, where is was expressed in a form valid for scattering of particles in general colour representations. The four-loop result, eq.~(\ref{eq:M42}), is entirely non-planar, a property that can be understood based on its direct connection to the Regge cut \cite{upcoming}, and it involves a purely adjoint quartic Casimir.

Finally, we have briefly discussed one application of these calculations, which exploit the fact that it is possible to apply infrared factorization to the scattering amplitude calculated in the high-energy limit, to extract the soft anomalous dimension within the same limit. In this context, the calculation of the even NLL amplitude discussed above made it possible to extract the soft anomalous dimension to all orders, which, at this logarithmic accuracy, is found to have an infinite radius of convergence. From the computation of the odd amplitude at NNLL accuracy we determine the soft anomalous dimension at this logarithmic accuracy to four loop. Ref.~\cite{Becher:2019avh} analyzed the colour structure of the soft anomalous dimension, incorporating the recently computed four-loop cusp anomalous dimension \cite{Henn:2019swt,vonManteuffel:2020vjv}, which introduces quartic Casimirs. Our results show that the soft anomalous dimension contains additional quartic Casimirs, beyond those associated with the cusp.

The knowledge of the soft anomalous dimension in the high-energy limit is extremely useful for bootstrap approaches, see for instance \cite{Almelid:2017qju}. In practice, the information obtained in particular kinematic limits, such as the Regge limit, can be used to reconstruct the structure of infrared divergences in full kinematic. Preliminary steps in this direction, using the new information obtained in \cite{Caron-Huot:2017zfo,Caron-Huot:2020grv,Falcioni:2020lvv}, have been carried out in \cite{upcoming}, and have been discussed in another talk in this conference, \cite{Maher:2021nlo}, to which we refer for further details.

\section*{Acknowledgements}

We would like to thank Simon Caron-Huot and Josha Reichel for stimulating collaborations on some of the topics discussed in this talk. 

\paragraph{Funding information}

EG, GF and NM are supported by the STFC Consolidated Grant ``Particle Physics at the Higgs Centre''. GF is supported by the ERC Starting Grant 715049 ``QCDforfuture'' with Prinipal Investigator Jennifer Smillie. CM's work is supported by the Italian
Ministry of University and Research (MIUR), grant PRIN 20172LNEEZ. LV is supported by Fellini, Fellowship for Innovation at INFN, funded by the European Union's Horizon 2020 research programme under the Marie Sk\l{}odowska-Curie Cofund Action, grant agreement no. 754496.


\bibliographystyle{SciPost_bibstyle} 
\bibliography{ReggeRefs.bib}

\nolinenumbers

\end{document}